\documentclass[acmlarge]{acmart}
\usepackage{array}
\usepackage{booktabs} % For formal tables
\usepackage{multirow} % For multirows
\usepackage{amssymb} % For the checkmark
\usepackage{graphicx} % For the rotation
\usepackage{enumitem}
\usepackage{subfig}

\newcommand{\score}{\texttt{ScoRe}}

\newcommand{\corerank}{\texttt{CoReRank}}
\newcommand{\fameforsale}{\texttt{Fame4Sale}}
\newcommand{\emscad}{\texttt{EMSCAD}}

\newcommand{\gsrank}{\texttt{GSRank}}
\newcommand{\speagle}{\texttt{SpEagle}}
\newcommand{\gsbp}{\texttt{GSBP}}
\newcommand{\revtwo}{\texttt{REV2}}
\newcommand{\ggspam}{\texttt{GGSpam}}
\newcommand{\defrauder}{\texttt{DeFrauder}}
\newcommand{\hawkeseye}{\texttt{HawkesEye}}
\newcommand{\mtlcollu}{\texttt{MTLCollu}}
\newcommand{\viewwgcca}{\texttt{View\_WGCCA}}
\newcommand{\scoreplus}{\texttt{ScoRe+}}

\AtBeginDocument{%
  \providecommand\BibTeX{{%
    \normalfont B\kern-0.5em{\scshape i\kern-0.25em b}\kern-0.8em\TeX}}}

\setcopyright{none}
\renewcommand\footnotetextcopyrightpermission[1]{}
\copyrightyear{2020}
\acmYear{2020}
\makeatletter
\def\@copyrightspace{\relax}
\makeatother
\begin{document}
%%
%% The "author" command and its associated commands are used to define
%% the authors and their affiliations.
%% Of note is the shared affiliation of the first two authors, and the
%% "authornote" and "authornotemark" commands
%% used to denote shared contribution to the research.
\title{Blackmarket-driven Collusion on Online Media: A Survey}
\author{Hridoy Sankar Dutta}
\authornote{This is the corresponding author}
\affiliation{%
  \institution{IIIT-Delhi, India}
  \city{New Delhi}
  \country{India}
}
\email{hridoyd@iiitd.ac.in}

% \author{Siva Charan Reddy Gangireddy}
% \affiliation{%
%   \institution{IIIT-Delhi, India}
%   \city{New Delhi}
%   \country{India}}
% \email{sivag@iiitd.ac.in}

\author{Tanmoy Chakraborty}
\affiliation{%
  \institution{IIIT-Delhi, India}
  \city{New Delhi}
  \country{India}}
\email{tanmoy@iiitd.ac.in}

%%
%% By default, the full list of authors will be used in the page
%% headers. Often, this list is too long, and will overlap
%% other information printed in the page headers. This command allows
%% the author to define a more concise list
%% of authors' names for this purpose.
\renewcommand{\shortauthors}{Dutta and Chakraborty}
%%
%% The abstract is a short summary of the work to be presented in the
%% article.
\begin{abstract}
  Online media platforms have enabled users to connect with individuals, organizations, and share their thoughts. Other than connectivity, these platforms also serve multiple purposes - education, promotion, updates, awareness, etc. Increasing the reputation of individuals in online media (\textit{aka Social growth}) is thus essential these days, particularly for business owners and event managers who are looking to improve their publicity and sales. The natural way of gaining social growth is a tedious task, which leads to the creation of unfair ways to boost the reputation of individuals artificially. Several online blackmarket services have developed thriving ecosystem with lucrative offers to attract content promoters for publicizing their content online. These services are operated in such a way that most of their inorganic activities are being unnoticed by the media authorities, and the customers of the blackmarket services are less likely to be spotted.  We refer to such unfair ways of bolstering social reputation in online media as \textit{collusion}. This survey is the first attempt to provide readers a comprehensive outline of the latest studies dealing with the identification and analysis of blackmarket-driven collusion in online media. We present a broad overview of the problem, definitions of the related problems and concepts, the taxonomy of the proposed approaches, description of the publicly available datasets and online tools, and discuss the outstanding issues. We believe that collusive entity detection is a newly emerging topic in anomaly detection and cyber-security research in general and the current survey will provide readers with an easy-to-access and comprehensive list of methods, tools and resources proposed so far for detecting and analyzing collusive entities on online media.
\end{abstract}
%%
%% Keywords. The author(s) should pick words that accurately describe
%% the work being presented. Separate the keywords with commas.
%%
%% This command processes the author and affiliation and title
%% information and builds the first part of the formatted document.
\maketitle
\section{Introduction}
The prosperity of online media has attracted people and organizations to join the platform and use it for several purposes - creating a network among commonalities, building and broadening their business, promoting/demoting e-commerce products, etc. This has led users to choose artificial ways of gaining social growth to get benefits within a short time. 
The main reason behind choosing artificial boosting is that the legitimate efforts of gaining appraisals (followers, retweets, likes, shares, etc.) take a significant amount of time and may not meet the actual needs of users. Such activities impose a significant threat to social media platforms as these are mostly against the Terms of Service of many social media platforms (see Section \ref{sec:terms_service} for more details). Such artificial boosting of social reputation/growth is often known as ``\textit{collusion}''  \cite{dutta2018retweet,Chetan:2019:CRD:3289600.3291010}.

\begin{figure}[!htbp]
\captionsetup[subfigure]{labelformat=empty}
    \subfloat[]{{\includegraphics[width=0.97\textwidth]{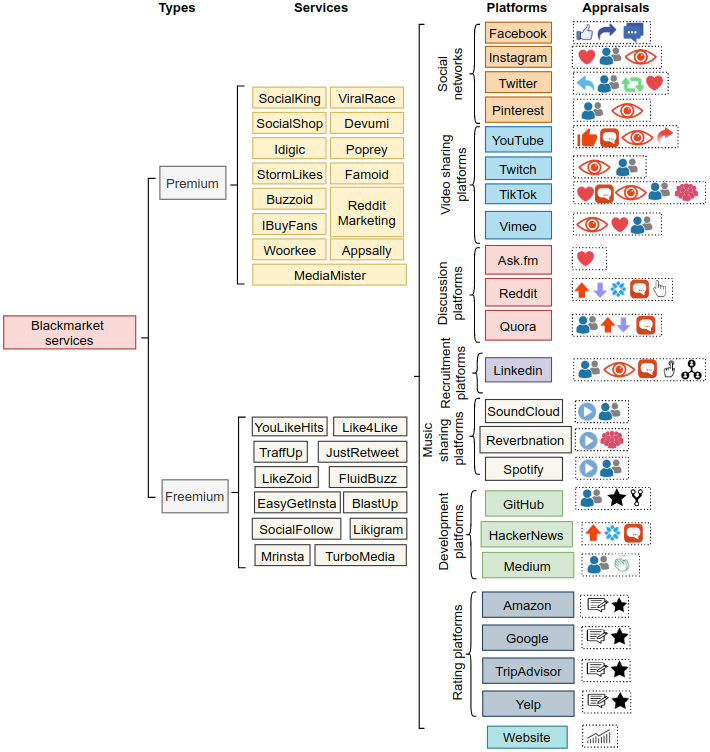} }} \\ \vspace{-9mm}
    \subfloat[]{{\includegraphics[width=\textwidth]{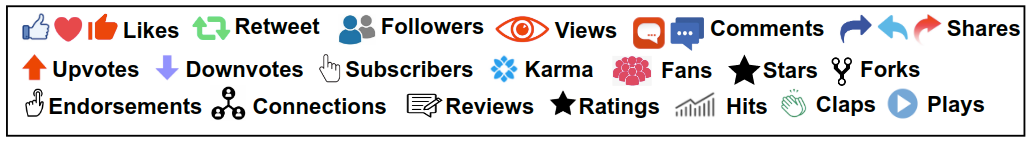} }}
    % % \caption{Example blackmarket website providing collusive appraisals to online media platforms such as Facebook, Instagram, Twitter, YouTube, SoundCloud, Spotify (name of the blackmarket redacted). }
    \vspace{-4mm}
    \caption{Illustration of blackmarket services with types, platforms, appraisals.}
    \label{fig:example_bm}
\end{figure}

According to a recent survey by HitSearch\footnote{\url{https://tinyurl.com/y39qtg2r}}, $98\%$ of content creators admitted to having spotted collusive followers among online influencers on Instagram. The adversarial impact of the collusive entities poses a massive threat to online media. These entities create an atmosphere where people start trusting their information due to the popularity they receive. For example, in the $2019$ UK general election, politicians approached the blackmarket services\footnote{\url{https://www.ics-digital.com/general-election-2019-mps-fake-twitter-followers/}} for online political campaigning in order to reach out to their potential voters. A study conducted by LawSuit\footnote{\url{https://lawsuit.org/politics-and-fake-social-media-followers/}}, a law management firm in the United States, reported that around $15\%$ of the Twitter users are non-human accounts driving more than $66\%$ links published to the site. It is also reported that most of the politicians currently in contention or conversation for the $2020$ US presidential election have a very high percentage or volume of non-human followers linked to their Twitter account. The above examples show how collusive entities boost the believability of information during events. Moreover, the limitation of humans' potential to distinguish between the collusive and genuine entities due to the vast amount of available information is an important concern that motivates to design methods for automatic identification of these entities. The collusive entities not only deceive people but also pollute the entire social space. Using the blackmarket services, the collusive entities can perform appraisals such as improving the credibility of rumors, propagate fake news, inflate/deflate the ratings of products in e-commerce platforms, gain popularity in video-sharing platforms, etc. Fig. \ref{fig:example_bm} illustrates how collusive entities gain artificial appraisals on various online media platforms. We request the reader to refer to Section \ref{sec:compromised} for a thorough explanation of Fig. \ref{fig:example_bm}. Recent studies that tackle the problem of collusive entity detection state that it is harder to discern these entities as they express a mixture of organic and inorganic activities \cite{dutta2018retweet}. Moreover, the problem is relatively new as compared to its related studies, e.g., bot/spam detection, fake account detection, etc. and recently has gained significant attention from diverse  research communities.

Although artificial boosting has become a regular practice with the increasing popularity of different online media platforms, there is a lack of coherent and collective attempts from different research communities to explore the micro-dynamics controlling such malpractice and the extent of its effect in manipulating social reputation. Most of the existing studies aim to detect  spam \cite{yardi2010detecting,wang2010don,thomas2011suspended,song2011spam,benevenuto2010detecting,mccord2011spam,thomas2013trafficking,chu2012detecting,zhang2012detecting,wang2013click,santos2014twitter,lin2013study,wang2015making}, fake \cite{gupta2013faking,gurajala2015fake,cook2014twitter,jin2013epidemiological,mehrotra2016detection,saez2014fake,buntain2017automatically,el2016fake,zhang2016discover,conti2012fakebook,gupta2014tweetcred,shen2014automatic,chen2015analysis} and bots \cite{yang2019arming,varol2017online,davis2016botornot,ferrara2016rise,subrahmanian2016darpa,chu2010tweeting,chu2012detecting,chavoshi2016debot,gilani2016stweeler,chavoshi2017temporal,dickerson2014using,morstatter2016new,cai2017behavior,kudugunta2018deep,zhang2011detecting,kantepe2017preprocessing}, and how these accounts are used for information propagation \cite{kupavskii2012prediction,bakshy2011everyone,lerman2010information,del2016spreading,vosoughi2018spread,shao2018spread,papanastasiou2018fake,cheng2014can,lerman2010information} in online media. Studies \cite{dutta2018retweet,dutta2019blackmarket} have shown that collusive users are neither fake users nor bots, but are normal human beings who express a mixture of organic and inorganic activities. Unlike bots, these users have no synchronicity across their behaviors \cite{dutta2018retweet}, which makes it difficult to design automated techniques to detect them.  In this paper, we present a comprehensive survey of existing literature on topics related to collusion in different online media platforms. Recently,  \citet{kumar2018false} surveyed three aspects of false information on the web and social media -- fake reviews, hoaxes, and false news. They also mentioned about the lack of publicly available datasets related to false information and social media rumors. In comparison to \cite{kumar2018false} and other related surveys, we mention our contributions in the remaining part of this section.
 %\cite{heymann2007fighting} surveyed solutions to detect spam on social websites.
%Though our work is closely related to the detection of the existing approaches/frameworks, we believe,

\subsection{Related Surveys and Our Contributions}
To our knowledge, \textit{this is the first survey to provide a detailed overview of collusive activities in online media}. The aim of this survey is to provide readers with a comprehensive overview of the major studies conducted in detecting and reasoning collusive activities across different online media platforms. The following four aspects make our  survey unique and different from other related surveys:
\begin{enumerate}[noitemsep,nolistsep]
    \item Existing surveys do not directly focus on the problem of ``collusive activities'' and are more centered around the detection and analysis of fake users, fraudsters and spammers on the web. Previous studies mentioned that unlike these problems, collusive activities are very different in nature~\cite{arora2020analyzing,dutta2018retweet,dutta2019blackmarket} due to the mixture of organic and inorganic activities.  In this paper, we conduct a thorough analysis of past studies dealing with collusive activities in different online platforms.
    \item Existing surveys mentioned only fraud and spam-related datasets. Here, we describe the datasets on collusion from multiple aspects - the type of dataset and the entities present in them.
    \item We also outline the annotation guidelines and evaluation metrics used for collusive entity detection, as mentioned in the related studies.
    \item We conclude the paper by highlighting a set of key challenges and open problems in the area of collusion in online media.
\end{enumerate}
We use the terms \textit{collusion} and \textit{artificial boosting of social growth} interchangeably throughout the paper. We refer to an \textit{action}/\textit{appraisal} as online media activity such as retweet, follow, review, view, subscription, etc., and an {\em entity} as a social entity such as user, tweet, post, review, etc. 

\subsection{Survey Methodology}

\subsubsection{Survey Scope} Since our scope is on investigating collusive entities in online media platforms, we will systematize the studies related to the analysis and detection of the collusive entities. As collusion is related to a few other aspects such as fake, bot, spam, etc., we partly focus on the previous works in those domains as well.

\subsubsection{Survey Organization}
In this article, we survey the existing algorithms, methodologies and applications in detecting and analyzing collusive activities in online media platforms. In Section \ref{sec:background}, we provide a broad overview and background of collusion in online media. We also analyze particular cases of collusion, show how collusion is closely related to other close concepts, see how collusion has been evolving and who are the main targets of it. Next, in Section \ref{sec:blackmarkets}, we provide some preliminary concepts on blackmarket services. In Section \ref{sec:compromised}, we show how collusion happens across multiple online media platforms and overview the state-of-the-art techniques for each platform. Section \ref{sec:collusion_types} explains two broad categories of collusive activities. To identify what has been done so far in the literature, in Section \ref{sec:collusion_progress}, we conduct a systematic literature review of the existing techniques. In Section \ref{sec:datasets}, we present the annotation guidelines, related datasets and evaluation metrics that can be used for studies in collusive entity detection. Being a under-explored research problem, there are plenty of important issues that need further attention. Pointers to such issues are mentioned in Section \ref{sec:future_opportunities}. Section \ref{sec:conclusion} concludes this survey with a summary of the main contributions and open problems.

\section{Background and Preliminaries} \label{sec:background}
\subsection{Overview of Online Media}

\subsubsection{Online Media  Platforms} 
\begin{figure*}[!t]
    \centering
    \includegraphics[width=\textwidth]{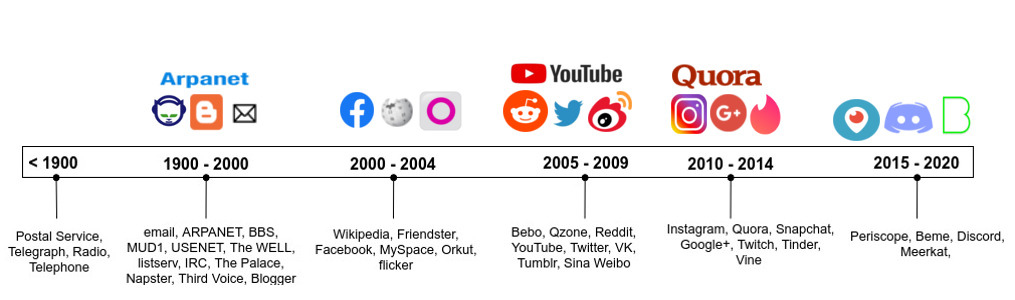}
    \caption{Evolution of online media platforms.}
    \label{fig:evolution}
\end{figure*}

Online media refers to the technologies present on the Internet that connect people or organizations to exchange information. The reason online media has become famous is the abundant sources of information offered by the Internet where a user can get access to the same news from several places. To keep users engaged, online media platforms also enable them to share their opinions on the information. In today's world, social media is considered to be a fast, inexpensive and effective way to reach a target audience. The history of online media started at the end of the 19th century with the arrival of Arpanet, email, blogging and bulletin boards. Prior to this, services such as telegram, radio, telephone, etc. were used to exchange information. However, their contributions were limited  as information exchange in these platforms did not take place ``online'' and was mostly used to send individual messages between two people. The rapid growth of the Internet in early 2000 set the real stage for the emergence of online media with the arrival of knowledge sharing and social networking sites. Fig. \ref{fig:evolution} depicts the history of online media starting from postal service in the early 90s to live-streaming (Periscope, Meerkat) and messaging platforms (Discord) in 2020. Example of online media platforms are social networking sites (e.g., Facebook, LinkedIn), microblogs (e.g., Twitter, Tumblr), wiki-based knowledge-sharing sites (e.g., Wikipedia), social news sites and websites of news media (e.g., Huffington Post), forums, mailing lists, newsgroups
community media sites (e.g., YouTube, Flickr, Instagram),
social Q \& A sites (e.g., Quora, Yahoo Answers), user reviews (e.g., Yelp, Amazon.com), social curation sites (e.g., Reddit, Pinterest) and location-based social networks (e.g., Foursquare).

\subsubsection{Terms of Service in Online Media Platforms} \label{sec:terms_service}
Terms of Service (ToS) in online media platforms refer to the rules and regulations to be agreed upon by the users of the services. Each service has its own policy against fake and spam engagements. For instance, Twitter has declared its own platform manipulation and spam policy\footnote{\url{https://help.twitter.com/en/rules-and-policies/platform-manipulation}}, YouTube has Fake Engagement Policy\footnote{\url{https://support.google.com/youtube/answer/3399767?hl=en}}, etc. 
These policies ensure that violations of the rules may result in permanent suspension of account and its content. The ToS followed by the online media platforms forbids artificially inflating own or others' appraisals (followers/retweets/views/likes/subscriptions). This includes selling or purchasing engagements using premium services, using or promoting third-party apps by posting content that helps in gaining engagements, trading to exchange engagements using freemium services, etc. Some of the terms proposed by Twitter on collusion-related activities are \textit{engagement churn} (first following a large number of unrelated Twitter accounts and then unfollowing them), \textit{indiscriminate engagement} (using third-party APIs or automated software to follow a large number of unrelated accounts in a short time period), \textit{aggressive engagement} (aggressively engaging with Tweets to drive traffic or attention to accounts).

% \subsection{Definition of collusion}
% \subsubsection{Definitions from Several Sources}
% We provide the basic definitions needed to understand the meaning of collusion from multiple aspects. 

% \begin{figure}[!t]
%     \centering
%     \includegraphics[width=0.5\linewidth]{images/collusion_definition.png}
%     \caption{Collusion: a general view.}
%     \label{fig:collusion_definition}
% \end{figure}

% \textbf{Collusion in Economics:} In economics, collusion is used to suppress rivalry, with an agreement among the competitors by relying on communication\cite{marshall2012economics}. In the case of industrial economics, collusion happens when firms get to an agreement on prices and quantities produced by each firm\cite{feuerstein2005collusion}. 

% \textbf{Collusion in Cryptography:} Collusion attack in cryptography is a phenomenon where a group of users combines their individual knowledge to gather unprotected content from any protected system\cite{doerr2005collusion}. For example, collusive users combine multiple digitally watermarked documents to create new documents with no underlying watermarks.

\subsection{Collusion in online media}
\subsubsection{Definition}
In general, collusion is defined as a covert and secret conspiracy or collaboration to deceive others. \textit{Collusion in online media} is a process by which users artificially gain social reputation, which violates the ToS of the online media platform. Online media entities involved in collusion approach blackmarket services to artificially inflate their social status. This results in entities to appear credible and legitimate to the end-users, thus leading to activities such as fake promotions, campaigns, misinformation, etc., thereby creating an inadequate social space. The blackmarket services provide eminent online media services ranging from online social networks to various other platforms such as rating/review platforms, video-sharing platforms and even recruitment platforms. In Section \ref{sec:compromised}, we discuss in detail how collusion happens in all these platforms. For example, a boost in YouTube views can transform a small event into a big campaign or a promotional event. Despite its apparent presence in the real world, collusion has remained as an underexplored concept. We go deeper into the notion of collusion by discussing particular cases and examples in the next subsection.

\subsubsection{Examples of Collusive Activities}
\begin{figure}[!t]
    \centering
    \subfloat[]{{\fbox{\includegraphics[width=0.51\columnwidth]{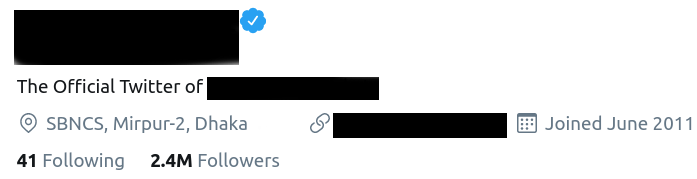} }}}%
    \hspace{0.5mm}
    \subfloat[]{{\fbox{\includegraphics[width=0.362\columnwidth]{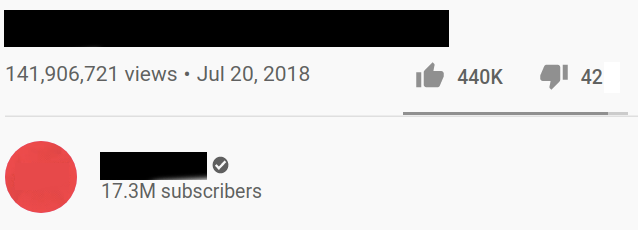} }}}%
    \caption{Example of collusive activities in online media. (a) an official Twitter account of an organization  registered in blackmarket services for collusive follower appraisals. (b) a video posted by a verified YouTube channel registered in blackmarket services for collusive like requests. Some profile information are blurred for the sake of anonymity.}
    \label{fig:example_collusion}
    \vspace{-5mm}
\end{figure}

In this section, we show two examples of how collusion happens in online media. Fig. \ref{fig:example_collusion} shows an example of collusive activities in online media. Note that we redacted the name of the Twitter user/YouTube channel to maintain anonymity. Fig. \ref{fig:example_collusion}(a) shows the official Twitter account of an organization registered in blackmarket services for collusive follower appraisals. Fig. \ref{fig:example_collusion}(b) shows a video posted by a verified YouTube channel registered in blackmarket services for collusive like requests. In both examples, the accounts are marked as \textit{verified} by Twitter/YouTube. The presence of verified online media entities in the blackmarket services clearly shows that the in-house algorithms deployed by these platforms have been unsuccessful in detecting such entities. This further motivates the problem and necessitates the development of automated techniques to detect these entities. In Section \ref{sec:social_networks}, we discuss in detail how online media users can request verification badge through the blackmarket services.

\subsection{Reasons behind Collusion}
In order to keep up with the pace of today's turbo-charged world, a large number of users like to go against the stream. Though shortcuts to success always conflict with the enormous efforts needed to be successful, it may also lead to some costly mistakes, eventually undermining the true goals. Online media is considered as the perfect convergence of communication and information. It has already become the most important source for public information, organizations and for individuals to promote their ideologies, products and services. Manipulating such an important source of information can result in a significant gain in terms of fame and finance for individuals/organizations. This is clearly evident in online social networking platforms like Twitter in the form of attaining fake followers, rating platforms like Amazon in the form of posting fake reviews, video streaming platforms like YouTube in the form of synthetic gain in viewership count and many more. With a huge competition in online media, it is hard for online media users who spend months creating new content but fail to reach their target audience. Blackmarket services serve as a complete social media toolkit that helps an online media entity to gain a stronger social media presence among its competitors. It is very effective for -- (i) companies who want to hype a campaign, (ii) musicians who want to promote new projects, album and releases, (iii) entertainment companies who want to publicize their shows, and (iv) online media moguls who need quick online media presence.  

\begin{table*}[!t]
\caption{Comparison between collusion and other related concepts.}\label{table:comparison_definition}
\centering
\begin{tabular}{l|p{60mm}|p{60mm}}
\hline
\textbf{Concept} & \textbf{Definition of the concept} & \textbf{Difference from collusion} \\\hline
Fake & Fake is to increase the visibility of others' content \cite{dutta2020hawkeseye}. & Collusion is self-focused i.e., to increase the visibility of their own content. \\ \hline
Bot & Bots have synchronous fraudulent activities \cite{giatsoglou2015nd}. & Collusive entities exhibit asynchronous fraudulent activities. \\\hline
Sockpuppetry & Sockpuppets are operated by a puppetmaster who  controls other user accounts \cite{kumar2017army}. & In collusion, every account is controlled by the real owner of the account. \\\hline
Malicious promotion & Promote a specific product/topic for a target audience \cite{li2014detecting}. & Collusion is more general and not necessarily focused on a specific product/topic.\\\hline
Spam & Consistently perform similar operations across multiple accounts to manipulate or undermine current trends \cite{gee2010twitter}. & Collusive entity may contain spammy contents, but not necessarily. \\\hline
Content polluters & Polluters post nearly identical contents, sometimes by randomly adding mentions to unrelated legitimate users \cite{lee2011seven}. & In collusion, content pollution never occurs as every action happens through the blackmarket services. \\\hline
Relationship Infiltrators & They follow the reciprocity in the relationship (follow/retweet/subscribe) to engage in spam activities \cite{lee2011seven}. & In collusion, reciprocity happens in freemium services to gain credits.\\ \hline
Slanderous users & They give a false
low rating with a positive review to confuse recommender systems \cite{xu2019slanderous} & Collusive users always give a high rating and positive review.\\
\hline
\end{tabular}\vspace{-5mm}
\end{table*}

\subsection{Challenges in  Collusion Detection}
Detecting collusive activities is a challenging task. \citet{dutta2018retweet} showed how collusive users show a mixed behavior of organic and inorganic activities. In the case of Twitter, these users, on the one hand, are  involved in retweeting (following) genuine tweets (genuine users); and on the other hand, they are involved in retweeting (following) tweets (users) submitted to the blackmarket services. This kind of mixed behavior makes it difficult to be detected by traditional fake user detection methods \cite{shah2017many}.
Moreover, collusive users are not bots; they are normal human beings. This makes it difficult to be flagged by bot detection methods   \cite{Chetan:2019:CRD:3289600.3291010}. A recent study by \citet{dutta2020detecting} investigated how collusion happens on YouTube. They designed web scrapers to collect a large set of YouTube videos and channels submitted to the blackmarket services for collusive appraisals. In the case of rating platforms like Amazon, methods to detect collusive reviewers have not been successful so far due to the lack of enough labeled data. The dynamic nature of the propagation of collusive activities is very complicated. Collusive activities can easily propagate and impact a large number of users in a short time by spreading misinformation. \citet{kim2018leveraging}
developed \texttt{CURB}, an online algorithm that leverages information from the crowd to prevent the spread of misinformation. They also reported that fact-checking organizations like \texttt{Snope} and \texttt{Politifact} are not able to properly limit the spread of misinformation as it requires significant human effort. Moreover, as collusion happens in multiple online media platforms, it is also a difficult task to understand the complete adversarial intent of the collusive entities. This raises concern on developing systems for early detection of collusive entities to limit its artificial social growth. Finally, due to the restrictions of the online media platforms on collecting the public data, the research community has very limited training data, which does not include all the necessary information of collusive entities. Most of the previous studies \cite{dutta2020detecting,arora2020analyzing} had to design their own scrapers due to the limitations and restrictions of the APIs provided by the online media platforms.

\subsection{Collusion and Other Related Concepts}
%Here, we compare collusion with several other related concepts. 
Some related concepts to collusion are fake, bot, sockpuppetry, malicious promotion, spam, content polluters, relationship infiltrators and slanderous users. We distinguish between these concepts and collusion in Table \ref{table:comparison_definition}. Even though these concepts differ from collusion, they are highly related. Therefore, studying the literature which focuses on these topics will give us a better knowledge of how to analyze and detect the collusive entities in online media platforms.

% \subsection{Evolution of collusion}

\section{A note on Blackmarket Services} \label{sec:blackmarkets}
Blackmarket services are the major controlling authorities of collusive activities. Customers join these services and contribute directly or indirectly to boost their online profiles artificially. The blackmarket services are divided into two types  based on the mode of service \cite{shah2017many} --
\begin{figure*}[!t]
    \centering
    \includegraphics[width=0.9\textwidth]{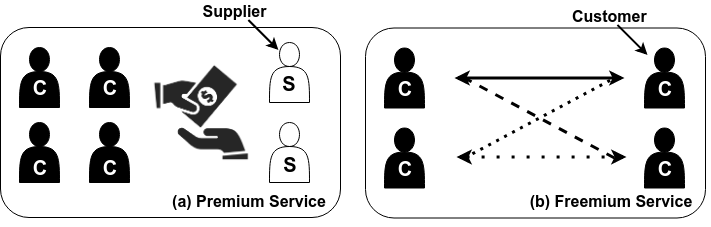}
    \caption{Illustration showing the working principle of premium and freemium services. }
    \label{fig:freemium_premium}
\end{figure*}
\textit{Premium} and \textit{Freemium}. \\
\textbf{1. Premium Services:} These services require the customers to pay a certain amount of money in order to obtain the facilities (e.g., SocialShop, RedSocial). Most of the premium services provide a comprehensive range of social media enhancement services for all purposes. These services also ensure strategic social media promotion to maintain an edge over the competitors, in the following ways: (i) location-specific actions, (ii) time during which users want to gain social growth, (iii) users gained from the services have eye-catching display pictures and filled out bios in the profile, and (iv) domain-specific actions (users gained from services can be from various domains such as fashion, music, blogging, etc.). These services offer actions in tiers (100, 1K, 10K, etc.) and lucrative offers to their customers and some additional facilities such as 100\% money-back guarantee, retention guarantee, complete privacy, etc. 
Few of these services ensure replenishment if the customer experiences any drop in the actions. We divide this type of services into two categories:
\begin{enumerate}
\item \textbf{ General 
Premium:} Here, customers have to choose a plan which suits their budget. 

\item \textbf{Auto Premium:} These services provide daily/weekly/monthly delivery service of actions. Customers need to select one of the auto action packages in advance for a time duration, e.g., 100 Twitter auto retweets (2-4 days). The main advantage of these services is to automate the process of social reputation to an extent. The basic principle of auto premium remains the same as that of general premium services but can be considered as a faster and effective way to boost social reputation. 
% Some of the services consider their auto service facility to maximize the extraordinary value and influence its traditional general premium service. 

\end{enumerate}
\noindent\textbf{2. Freemium Services:} These services are free to use; but they also have premium subscription plans (e.g., YouLikeHits, Like4Like). The main idea is to get customers familiar with the workflow of the services and motivate them to opt for the premium plans. Freemium services operate in one of the three ways: \textit{social-share services, auto-time freemium services, and credit-based services}. More details on these types can be found in  \cite{dutta2018retweet} and \cite{dutta2019blackmarket}.
Freemium services operate in one of the three ways \cite{dutta2018retweet}:
\begin{enumerate}
    \item  \textbf{ Social-share Services:} These services require customers to perform social actions on multiple platforms in order to get appraisals for their content. Some of the possible actions are share/like/follow on Facebook,  follow/like/view/comment on Instagram, like/view/share on YouTube, etc. 
\item \textbf{Auto-time Freemium services:} These services require the customers to get access tokens from the services, after which they can request for a fixed number of actions for a time duration, e.g., 10-50 retweets in 10 minutes window.
\item \textbf{Credit-based Services}: These services are operated based on a `give and take' relationship. Customers of these services lose credits when other customers perform actions on their submitted content. Similarly, customers gain credits when they perform actions on the content of other customers.
\end{enumerate}

\begin{figure}[!htbp]
    \centering
    \subfloat[]{{\fbox{\includegraphics[width=0.435\columnwidth]{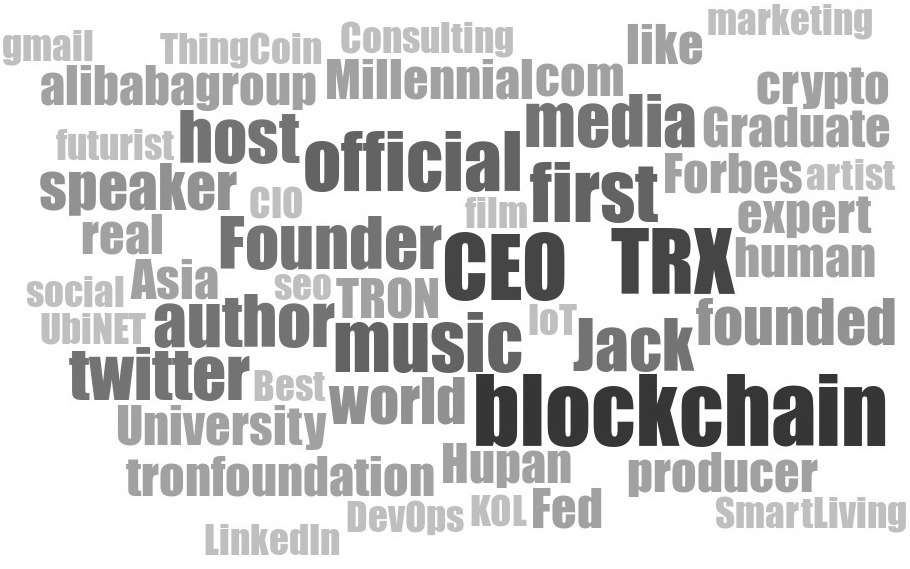} }}}%
    \hspace{0.2mm}
    \subfloat[]{{\fbox{\includegraphics[width=0.47\columnwidth]{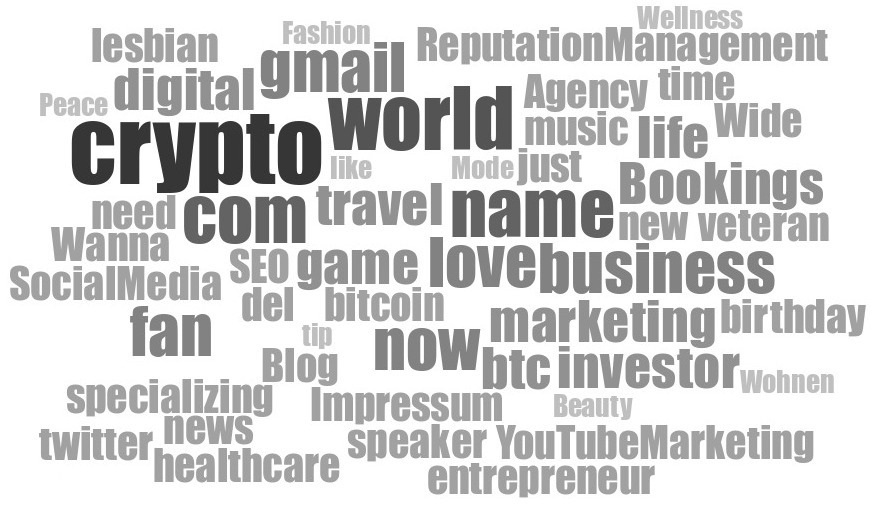} }}}%
    \caption{Wordclouds generated from the profile description of customers in (a) premium and (b) freemium services. Reprinted with permission from \cite{dutta2019blackmarket}. }
    \label{fig:example_wordcloud}
\end{figure}

Fig. \ref{fig:freemium_premium}(a) shows the working of premium services. Here, two types of entities are involved -- \textit{customers} that ask for appraisals and \textit{suppliers} those supply appraisals. Fig. \ref{fig:freemium_premium}(b) shows the working of freemium services. Here, \textit{customers} are involved in a credit-based ecosystem; hence \textit{customers} are also the \textit{suppliers}. The wordclouds of the description/bios or premium and freemium users observed by \citet{dutta2019blackmarket} are shown in Fig. \ref{fig:example_wordcloud}.

\section{Compromised Online Platforms} \label{sec:compromised}
As discussed in previous sections, collusion happens across multiple online media platforms. In this section, we discuss in detail how appraisals in these platforms are artificially manipulated by the collusive entities. We will look into seven types of platforms: \textit{Social networks}, \textit{rating/review platforms}, \textit{video streaming platforms}, \textit{recruitment platforms}, \textit{discussion platforms}, \textit{music sharing platforms} and \textit{development platforms}. Other than the above-mentioned platforms, we will also look into the artificial manipulation of website traffic through blackmarket services. Fig. \ref{fig:example_bm} illustrates how different types of blackmarket services provide collusive appraisals to various online media platforms.

\subsection{Social Networks} \label{sec:social_networks}

Social networks serve as a platform to build social relations among users of common interests (personal or professional). Platforms like Twitter, Facebook and Instagram allow their users to perform actions such as post, like, retweet, follow, share, etc. Nowadays, these platforms also serve as real-time news delivery services and a medium for the business owners to connect with their customers and thus expand their outreach. However, the natural way to attract users  is usually a tedious task and takes significant time. This motivates users to choose an artificial way of gaining appraisals.

\textit{Facebook.} Facebook is the most popular social network to connect and share with family and friends online. Facebook has four types of appraisals: \textit{shares}, \textit{likes}, \textit{comments}, \textit{followers}. Large number of appraisals on Facebook acts as a form of social proof for the sign of popularity and importance.

\textit{Twitter.} Twitter is a microblogging service where users write tweets about topics such as politics, sport, cooking, fashion, etc.  Twitter has three types of appraisals: \textit{retweets}, \textit{likes} and \textit{followers}. Acquiring more appraisals  on Twitter helps to increase the user's social signals and attract more visitors to the profile. 

\textit{Instagram.} Instagram is a social networking platform that enables users to share images or videos with their audience. Users can upload photos, videos and share them with their followers or with a group of friends. Instagram has three types of appraisals: \textit{likes}, \textit{followers} and \textit{views}. Higher appraisals are the key to visibility on Instagram. The more appraisals a post receives on Instagram, the higher is the post rank in search results and on the Explore page. 

\textit{Pinterest.}: Pinterest is an image-sharing  platform that is designed as a visual discovery engine for finding ideas like recipes, home and style inspiration, etc. A message on Pinterest is known as Pin. Pinterest has two types of appraisals: \textit{likes} and \textit{followers}.

Other than the common appraisals, we found few blackmarket services where social networking users can request for a verification badge. 
\texttt{Verifiedbadge}\footnote{\url{https://verifiedbadge.co/}}, \texttt{StaticKing}\footnote{\url{https://www.staticking.com/}}, \texttt{Prime badges}\footnote{\url{https://primebadges.com/}} and \texttt{SocialKing}\footnote{\url{https://www.socialking.in/}}  are the most popular platforms providing such appraisals. Verification badges are coveted checkmark badges that enhance the user's social media presence and improve credibility on the platform. The minimum requirements to request for verification badges through blackmarket services are as follows: 
\begin{itemize}
\item The user/organization should be a celebrity, journalist, popular brand, government official or sports company.
\item The user/organization should have a Wikipedia page or media coverage in leading online news portals.
\item For social media platforms where subscription is an appraisal, the user should have a minimum of 100K subscribers.
\end{itemize}

There is a vast literature on the detection and analysis of collusive entities in social networks. \citet{shah2017many} and \citet{dutta2018retweet} are two of the first few studies that investigate collusive entities on Twitter registered in blackmarket services. \citet{dutta2018retweet} trained multiple state-of-the-art supervised classifiers using a set of 64 features to distinguish collusive users from genuine users. They also divided the set of collusive users into three categories: \textit{bots}, \textit{promotional customers} and \textit{normal customers}. \citet{arora2020analyzing} obtained better classification performance than \citet{dutta2018retweet} by incorporating the content-level and network-level properties of Twitter users in a multi-task setting. \citet{de2014paying} performed the first work on Facebook where the authors presented a comparative measurement study of page promotion methods. \citet{sen2018worth} conducted the first work on Instagram where they developed an automated mechanism to detect fake likes on Instagram. We discuss in detail about these studies in Section \ref{sec:collusion_progress}. 

% Overall, the major aim of these approaches is to assist social networking platforms in identifying collusive frauds. All the algorithms perform pretty well on their respective datasets, with a reported accuracy of 70-80\%.

\subsection{Rating/Review Platforms}
Rating/Review platforms allow users to rate or share their opinion about entities, e.g.,  products, applications, food items, restaurants, movies, etc. Examples of these platforms are e-commerce platforms (e.g., Amazon), travel platforms (e.g., TripAdvisor), business rating platforms (e.g., Yelp, Google review), etc.

\textit{Amazon.} Amazon is the world's largest e-commerce platform and is considered the ultimate hub for selling merchandise on the web. It allows two types of appraisals: \textit{reviews} and \textit{ratings}. High ratings and reviews for a product on Amazon attract customers and make the product more trustworthy.

\textit{Google.} Google provides valuable information to businesses and its customers using two types of appraisals: \textit{reviews} and \textit{ratings}. Higher ratings and reviews on Google help to improve the business and enhance local search rankings.

\textit{TripAdvisor.} TripAdvisor is an online travel platform that offers multiple services such as online hotel reservations, travel experiences, restaurant reviews, etc. Similar to other  platforms, TripAdvisor has two types of appraisals: \textit{reviews} and \textit{ratings}. Positive reviews from previous occupants help the business to improve its reputation and increase the customer base.

\textit{Yelp.} Yelp is a local business review platform that collects crowdsourced data from its users. Yelp has two types of appraisals: \textit{reviews} and \textit{ratings}. Getting more reviews in turn improves business reputation and gets a lot more customers with free traffic.

Nowadays,  online reviews/ratings are becoming highly relevant for customers to take any purchase-related decisions. As these appraisals play a significant role in deciding the sentiment/popularity of a product/business, there is a massive scope for collusion among sellers/buyers to manipulate it artificially. \cite{jindal2010finding,li2011learning,wang2011review} are a few initial studies on detecting fake reviews in review patterns representing unusual behaviors of reviewers. Another early work by \citet{li2014spotting} identified fake reviews on Dianping, the largest Chinese review hosting site. The authors proposed a supervised learning algorithm to identify fake reviews in a heterogeneous network of users, reviews and IP addresses. \citet{mukherjee2012spotting} performed one of the first attempts  to detect fraudulent reviewer groups in e-commerce platforms. Recently, \citet{kumar2018rev2} identified users in rating platforms who give fraudulent ratings for excessive monetary gains. Most of the studies in review platforms are focused on e-commerce platforms. We believe the newer platforms such as Google reviews, Yelp and TripAdvisor would open more research questions as these platforms contain reviews of millions of hotels, restaurants, attractions and other tourist-related businesses.

%As the magnitude of this problem is huge, there is a need for it to be addressed, and this survey covers some of the important works that have been done to tackle this problem. 

%They reported some unfair users to the review fraud investigation team in Flipkart (an Indian e-commerce company), and 85\% of them were marked, fraudulent users. 

\subsection{Video Streaming Platforms}
Video streaming platforms are mostly used for sharing videos and streaming live videos. These platforms allow users to upload, view, rate, share, add to playlists, report, comment on videos and subscribe to other users. Here, we discuss how appraisals on video streaming platforms are inflated artificially. With the increase in the popularity of live streaming came the concept of ``astroturfing'' -- a broader and sophisticated term referring to the synthetic increase of appraisals in an online social network by means of blackmarket services. The consequences of such synthetic inflation are not only restricted to increase monetary benefits, directory listings and partnership benefits, but also expanded to better recommendation rankings thus doctoring the experiences of viewers who are recommended all these boosted materials instead of genuine materials produced by the honest broadcasters.

\textit{YouTube.} YouTube is a video sharing platform where users can create their own profile, upload videos, watch, like and comment on other videos. YouTube has four types of appraisals: \textit{likes}, \textit{comments}, \textit{subscribers}, \textit{views}. Likes, comments and views are for the videos and subscribers are for the channels. Higher the number of YouTube users interacts with a video/channel, higher that video/channel will be listed to other users. These appraisals are considered as the measure of engagement and not only make the entity popular but also help the channel with sponsorship opportunities and monetization options.

\textit{Twitch.} Twitch is the most popular live streaming platform on the web. It is mostly used by gamers to stream their games while other users can watch them. Twitch has two types of appraisals: \textit{followers} and \textit{views}. The higher views a channel has, the higher is its popularity on Twitch and to be ranked on the featured list. Twitch is usually considered to be one of the most difficult platforms to earn quick popularity due to the presence of so many streamers using the platform. 

\textit{Tiktok.} Tiktok is a video sharing social network which is primarily used to create short videos of 3 to 15 seconds. Tiktok has two types of appraisals: \textit{likes} and \textit{followers}. Getting more likes on Tiktok posts increases the chance to maintain its presence and make it famous amongst the audience.

\textit{Vimeo.} Vimeo is a video hosting and sharing platform that allows users to upload and promote their videos with a high degree of customization which is not available on other competing platforms. Vimeo has two types of appraisals: \textit{likes} and \textit{followers}. A higher number of appraisals helps to show up the video in the suggested videos list created by Vimeo's algorithm. 

 Few studies addressed issues of astroturfing in video streaming platforms. \citet{shah2017flock} made the pioneering attempt to combat astroturfing on live-streaming platforms. He proposed \texttt{Flock}, an unsupervised method to identify botted broadcasts and their constituent botted views. Note that he didn't disclose the name of the live streaming corporation on which the study was performed. In a recent study, \citet{dutta2020detecting} proposed \texttt{CollATe}, an unsupervised method to detect collusive entities on YouTube. The method utilizes the metadata features, temporal features and textual features of a video to detect whether it is collusive or not.

\subsection{Recruitment Platforms}
Recruitment platforms are employment-oriented platforms that also provide professional networking among users. These platforms offer the opportunity to discover new professionals either locally or internationally and help one with their professional endeavors. Hiring new employees is a crucial part of every organization, which starts with posting new job ads and ends with recruitment. It can be thought of as a multi-step process, which is normally very time-consuming and prone to human errors. With the advent and rise of automated systems, the hiring process of an organization is being done in the cloud with the help of tools such as  \textit{Applicant-Tracking-System (ATS)}\footnote{\url{https://en.wikipedia.org/wiki/Applicant_tracking_system}}. ATS makes the hiring process faster and accurate by preparing job ads, posting them online, collecting applicant resumes, making efficient communication with them and finding the best fit resumes for the organization. However, increasing use of ATS also invokes various disadvantages such as spammers compromising the job seekers' privacy, slandering the reputation of the organizations and financially hurting it by manipulating the normal flow of functioning of the system - most frequently the job ads publishing process (often recognized as the employment scam).

\textit{LinkedIn.} LinkedIn is the most popular online recruitment platform where employers can post jobs, and job seekers can post their profiles. There are four types of appraisals on LinkedIn: \textit{followers}, \textit{recommendations}, \textit{endorsements}, and \textit{connections}. A higher number of connections and followers on LinkedIn helps the user to gain attention. LinkedIn endorsements help to add validity to the user's profile by backing up his/her work experience.

\citet{adikari2020identifying} identified fake profiles on LinkedIn. The authors considered state-of-the-art supervised classifiers designed on a set of profile-based features to detect fake profiles.  \citet{prieto2013detecting} detected spammers on LinkedIn based on a set of heuristics and their combinations using a supervised classifier. Another work by \citet{vidros2017automatic} tackled the problem of Online Recruitment Frauds (ORF) (see Sec. \ref{sec:collusion_progress} for more details). The authors also released a publicly available dataset of $17,880$ annotated job ads ($17,014$ legitimate and $866$ fraudulent job ads) from various recruitment platforms.

\subsection{Discussion Platforms}
Discussion platforms are content sharing platforms primarily used to entertain community-based discussions. The discussions can be in the form of question-and-answers or content that other users have submitted using links, text posts or images. Here, we explain how collusion happens on discussion platforms:

\textit{Quora.} Quora is a question-and-answer based discussion forum that empowers the user to ask questions on any subject and connect with like-minded people who contribute unique insights and high-quality answers. Quora has four types of appraisals: \textit{followers}, \textit{upvotes}, \textit{downvotes} and \textit{comments}. Higher the number of followers and upvotes a user gains for his/her answers, higher is the ranking factor and the more the answers are displayed to other users. The aim of a user on Quora is to be named as a top writer in the long run. Answers written by top writers on Quora are considered as expert opinions and are also displayed in the featured list.

\textit{ASKfm.} ASKfm is another question-and-answers based discussion forum and is mostly used by users to post questions anonymously. The platform has two types of appraisals: \textit{followers} and \textit{likes}. Higher likes on answers grow its rating on ASKfm at a faster rate.

\textit{Reddit.} Reddit is a social news-based discussion forum that allows users to discuss and vote on content that other users have submitted. Reddit has five types of appraisals: \textit{subscribers}, \textit{upvotes}, \textit{downvotes}, \textit{karma} and \textit{comments}. Having more karma on Reddit allows the user to post more often on the platform and gives him/her more reputation. More number of upvotes on posts helps users gain more exposure, which eventually pushes the posts higher up on the targeted Reddit or Subreddit.

Studies in discussion platforms have been conducted with the goal of manipulating the visibility of political threads on Reddit \cite{carman2018manipulating}.  The authors measured the effect of manipulation of upvotes and downvotes on article visibility and user engagement by comparing Reddit threads whose visibility is artificially increased. Another work by \citet{shen2019discourse} investigated polarized user responses on an update to Reddit's quarantine policy.

\subsection{Music Sharing Platforms}
Music sharing platforms enable users to upload, promote and share audio. Here, we explain how collusion happens on music sharing platforms.

\textit{Soundcloud.} Soundcloud is an audio sharing platform that connects the community of music creators, listeners and curators. Soundcloud has three types of appraisals: \textit{plays}, \textit{followers} and \textit{likes}. A higher number of followers and likes creates a massive fan base for creators and gets more attention from the community. The popularity of a soundtrack in the platform is driven by the number of plays it receives. Plays attract users to hear the music as they feel to check something which is liked by most of the other users. 

\textit{Reverbnation.} Reverbnation is an independent music sharing platform where musicians, producers and venues collaborate with each other. The platform has two types of appraisals: \textit{plays} and \textit{fans}. Higher the number of plays in an audio or video, higher is the rank of the artist on the platform. 

Some studies have been conducted on fraudulent entity detection in music sharing platforms. \citet{bruns2018detecting} investigated  Twitter bots that help in promoting SoundCloud tracks. The authors also proposed a number of social media metrics that help to identify bot-like behavior in the sharing of such content. Another work by \cite{ross2018social} proposed a method to distinguish between two groups of SoundCloud accounts -- bots and humans.

\subsection{Development Platforms}
Interestingly, we found a few popular development platforms where collusion happens. 

\textit{GitHub.} GitHub is a repository hosting service that provides distributed version control and source code management (SCM) functionality. GitHub has three types of appraisals: \textit{followers}, \textit{stars} and \textit{forks}. User profiles with more followers make the account more popular. Similarly, stars and forks are the metrics to show the popularity of a repository. Blackmarket services help to deliver GitHub followers, stars and forks from real and active people. 

\textit{Hackernews.} Hackernews is the most popular discussion platform for developers. It has three types of appraisals: \textit{upvotes}, \textit{karma} and \textit{comments}. Higher the number of comments and upvotes on a post, higher is its popularity. User profiles with high karma can perform additional appraisals on a post: \textit{downvote}, \textit{making polls} etc. Blackmarket services help to deliver upvotes and karma on Hackernews simply by adding the post link and making the payment. 

\textit{Medium.} Medium is an online publishing platform that is commonly used by developers to share ideas, knowledge and perspectives. It has two types of appraisals: \textit{followers}
and \textit{claps}. A higher number of claps in a post ranks it higher in feed and search results. Similarly, a higher number of followers increases the post's reach and view. Blackmarket services help to deliver followers and claps by adding the post link and making the payment.

Most of the studies in development platforms are on measuring user influence and identifying unusual commit behavior by analyzing the attributes of the platforms. \citet{hu2018user} measured the user influence on GitHub using the network structure of the following relation, star relation, fork relation and user activities. The authors also introduced an author-level H-index (also known as H-factor) to measure users' capability. \citet{goyal2018identifying} identified the unusual changes in the commit history of GitHub repositories. The authors designed an anomaly detection model based on commit characteristics of a repository. 
% To facilitate future research on examining and detecting collusion in development platforms, we have compiled a comprehensive list of the blackmarket services and the appraisals on each platform (c.f. Fig. \ref{fig:example_bm}). 
One potential research direction is to examine how collusive appraisals on these platforms help to popularize the reputation of the repositories/users. Another potential research direction is to develop collusive entity detection techniques by considering the hidden relations between the entities of the platform.

\subsection{Other Platforms}
Other than the online media, artificial boosting is also observed in other platforms. Website owners can avail premium/freemium blackmarket services to get traffic on their website. The idea is to endow the responsibility for the necessary amount of web traffic to some other company. In blackmarket terms, the appraisal is called as \textit{``hits''} or \textit{``web traffic''}. Gaining artificial hits helps the website to gain popularity faster compared to the slow process of working on growing organic visitors via Search Engine Optimization. Blackmarket services offer two types of web traffic: \textit{Regional traffic} where the customers can opt for traffic from a specific country or region and \textit{Niche traffic} where customers can opt for targeted traffic that focuses on a specific type of business such as music-based, e-commerce based, etc. To activate traffic to a website, customers have to enter the URL of the website, the number of visitors required and preferred timespan for delivery. 

% Most of the works on fraudulent activities in websites are based on using ads to monetize and popularize the site. Few studies have been conducted on techniques to maintain the reputation and popularity of websites. 
% We believe that this area of research has tremendous potential in 

% \subsection{Working of blackmarket services}
\section{Types of Collusive Activities} \label{sec:collusion_types}
Collusive activities can be categorized based on the mode of collaboration - \textit{individual collusion} and \textit{group collusion}. In this section, we will discuss the two types of collusion and how the individual collusion differs from group collusion when providing collusive appraisals.

\subsection{Individual Collusion} 
Individual collusion happens when the collusive activities of individuals are independent of each other; however, they are guided by centralized blackmarket authorities. Understanding individual collusion has been studied to some extent in the literature. Most of the existing studies used supervised models based on behavioral features and profile features \cite{dutta2018retweet}. Some network-based approaches infer anomaly scores for nodes/edges in the network (tweet-user network, product-user network, etc.) and rank them to spot suspicious users \cite{wang2011review,kumar2018rev2,Chetan:2019:CRD:3289600.3291010,shah2016edgecentric,hooi2016birdnest}. Existing studies mostly focused on the behavioral dynamics of individuals. However, these approaches fail when it comes to detecting group collusion. Group collusion can be more damaging as they can take the total control of the appraisals for an entity due to its size. 

\subsection{Group Collusion}
Group-level collusion takes place when a set of individuals collaborate as a group to perform collusive activities. Such collective behavior is, therefore, more subtle than individual behavior. At an individual level, activities might be normal; however, at the group level, they might be substantially different from the normal behavior. Moreover, it may not be possible to understand the actual dynamics of a group by aggregating the behavior of its members due to the complicated, multi-faceted and evolving nature of inter-personal dynamics \cite{dhawan2017spotting}. Some properties of collusive groups are as follows -- (i) members in collusive group work in shorter time frames and create maximum impact (e.g., flooding deceptive opinions), (ii) members of the group may or may not know each other (agreement by a contracting agency), (iii) multiple accounts within a group can be controlled by a single master account ({existence of Sockpuppets} \cite{KumarCLS17}), (iv) larger the size of the group, the more damaging the group is, and (v) group members have a high chance of performing similar activities ({group members posting similar reviews or writing same comments}). 
%In group-level collusion, when each member of the group is considered separately, they might not look suspicious. However, when considered as a group, their fraudulent behavior might be disclosed. This is an important research problem in this area. 
Past studies in this direction mostly detected groups using Frequent Itemset Mining (FIM) and ranked groups based on different group-level spam indicators \cite{wang2018graph,gupta2019malreg}. 
% Thus, they focused more on ranking collusive groups, paying less attention to judge the quality of the detected groups, which still remains an unsolved problem. 
\citet{wang2016detecting} pointed out several limitations of FIM for group detection -- high computational complexity at low minimum support, absence of temporal information, unable to capture overlapping groups, prone to detect small and tighter groups, etc. \citet{liu2017holoscope} proposed \texttt{HoloScope} to find a group of suspicious users in rating platforms based on the contrasting behavior of  fraudsters and honest users in terms of topology, temporal spikes and rating deviation.

\section{Progress in collusive entity detection} \label{sec:collusion_progress}
Despite the increasing interest in analyzing and designing anomalous activities on the web, there has been limited work in collusive entity detection. Most of the previous studies in collusive entity detection are limited to social networks and rating platforms. In this section, we provide a summary of the reviewed papers to summarize the central idea and provide the basics of the models. To structure the descriptions, the relevant papers are grouped by the type of approaches  proposed in the paper. We categorize the methodologies followed by the previous studies into one of the following types: (i) feature based, (ii) graph based, and (iii) deep learning based. 
% Table \ref{table:comparison_approaches} presents a summary of the existing studies. \\

\subsection{Feature Based Detection} The majority of the papers model the behavioral properties of the users in online media platforms. Feature-based methods can be used to distinguish between collusive entities from genuine entities. The aim is to design a set of features that can (well) represent and capture various behavioral characteristics of the entities. Recently, researchers have also focused on the linguistic behavior of the collusive entities, such as deceptive information \cite{alowibdi2015deception, giatsoglou2015retweeting}, lexical analysis \cite{inuwa2018lexical} and sentiment analysis \cite{dickerson2014using}. 

\citet{dutta2018retweet} reported that collusive users show an amalgamation of organic and inorganic activities and there is no synchronicity among their behaviors. They substantiated their finding by comparing collusive retweeting behavior with that of normal retweet fraudsters. Here, normal retweet fraudsters are those retweeters who are abnormally synchronized in some patterns. The authors then collected data from four freemium blackmarket services (credit-based services). Three human annotators were asked to label the blackmarket customers into bots, promotional customers and normal customers based on a set of annotation rules. The data is used to tackle two problems - (i) a four-class classification problem (genuine, bot, promotional, normal), and (ii) a binary classification problem (genuine and combining all types of customers into a single class). In order to detect collusive retweeters, the authors developed \score, a supervised model to detect collusive retweeters based on five sets of features: profile features, social network features, user activity features, likelihood features and fluctuation features. The authors also developed a chrome browser extension to detect the collusive retweeters in real-time. To study the underlying behavior of the blackmarket services, \citet{dutta2019blackmarket}  extended their previous work \cite{dutta2018retweet} to provide an in-depth analysis of collusive users involved in premium and freemium blackmarket services. The authors collected the dataset from four premium and four freemium blackmarket services. They analyzed the activities of collusive users based on retweet-centric, network-centric, profile-centric and timeline-centric properties. They showed that unlike premium services, the credit-system in freemium services is the primary reason behind the unusual functioning of freemium collusive users. They also developed an updated version of the chrome extension by incorporating the features calculated from both premium and freemium users. Fig. \ref{fig:example_wordcloud} shows the wordcloud of the text present in the description of premium and freemium retweeters collected from the blackmarket services. It can be seen that premium retweeters are associated with high profile accounts having keywords such as `CEO', `official', `speaker', `founder'. However, freemium retweeters are associated with advertising agencies having keywords such as `like', `agency', `SocialMedia' and `YouTubeMarketing'.
        
    \citet{dutta2020hawkeseye} proposed \texttt{HawkesEye}, a classifier based on the Hawkes process and topic modeling to detect collusive retweeters on Twitter. They collected tweets  using three hashtags: \#deletefacebook, \#cambridgeanalytica, \#facebookdataleaks and manually annotated the retweeters as ``fake'' or ``genuine''. The \texttt{HawkesEye} model takes the temporal information of retweet objects of a user and topical information of the retweet texts as input. The temporal information is modeled using the Hawkes process due to its self-excitation property, and the topical information is modeled using Latent Dirichlet allocation (LDA). The authors reported highly accurate results on the classification task. Comparisons were made with respect to well-known bot detection approaches \cite{chavoshi2016debot} and collusive retweeter detection approaches \cite{dutta2018retweet}. 

\begin{figure*}[!t]
\minipage{0.33\textwidth}
  \includegraphics[width=\linewidth]{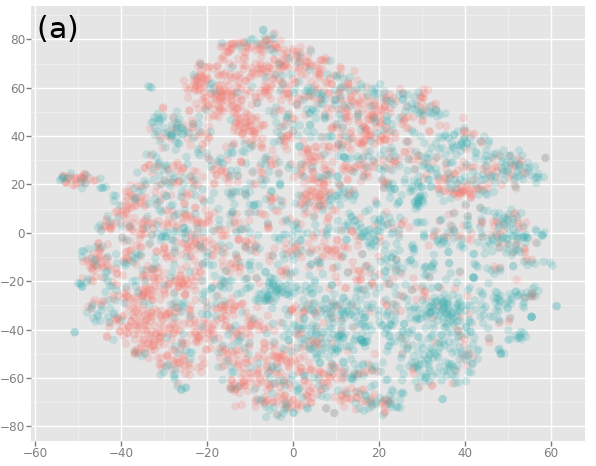}
%   \caption{(a) Tweet2Vec}
%   \label{fig:awesome_image3}
\endminipage
\minipage{0.33\textwidth}%
  \includegraphics[width=\linewidth]{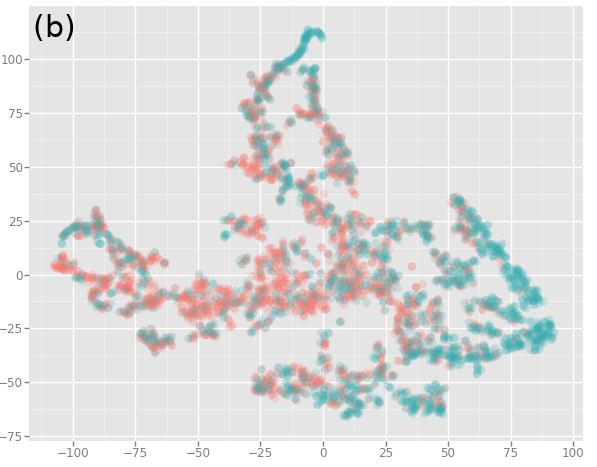}
%   \caption{(b) SCoRe}
%   \label{fig:awesome_image2}
\endminipage
\minipage{0.33\textwidth}
  \includegraphics[width=\linewidth]{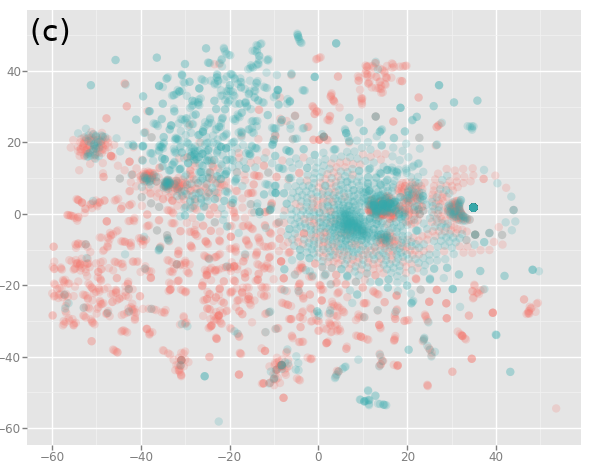}
%   \caption{(c) Retweet network}
%   \label{fig:awesome_image1}
\endminipage\hfill
\minipage{0.33\textwidth}
  \includegraphics[width=\linewidth]{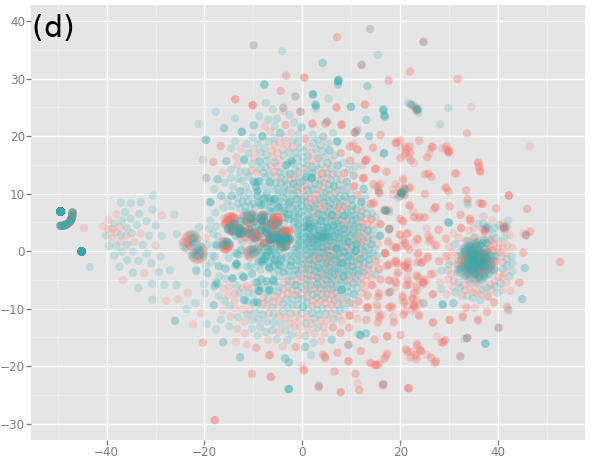}
%   \caption{(d) Quote network}
%   \label{fig:awesome_image3}
\endminipage
\minipage{0.33\textwidth}
  \includegraphics[width=\linewidth]{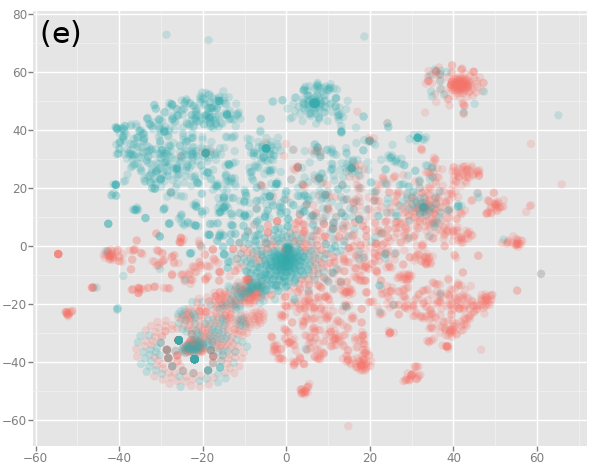}
%   \caption{(e) Follower network}
%   \label{fig:awesome_image3}
\endminipage
\minipage{0.33\textwidth}%
  \includegraphics[width=\linewidth]{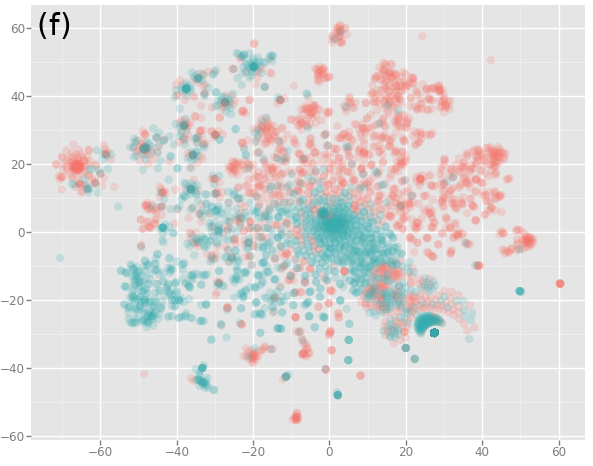}
%   \caption{(f) Followee network}
%   \label{fig:awesome_image2}
\endminipage\hfill
\caption{t-SNE visualization of representations of collusive (red) and genuine (green) users created using (a) Tweet2Vec, (b) \texttt{SCoRe}\cite{dutta2018retweet}, (c) Retweet network, (d) Quote network, (e) Follower network, (f) Followee network. Reprinted with permission from \citet{arora2020analyzing}.}
\label{fig:parameters}
%\vspace{-5mm}
\end{figure*}

    More recently, \citet{arora2020analyzing} improved the collusive retweeter detection task by considering the multifaceted characteristics of a collusive user. They created multiple views for a user by examining representations from the content, attributes and social network. They then proposed a multiview learning-based approach based on Weighted Generalized Canonical Correlation Analysis (WGCCA) to combine individual representations (views) of a user to derive the final user embeddings. The authors reported that as each view is more and less helpful for the classification experiment (c.f. t-SNE visualizations in fig. \ref{fig:parameters}), they assigned a weight to each view differently instead of treating each view equally.

   Several research studies have been conducted about the
automatic detection of collusive followers on Twitter. In an earlier study, \citet{liu2016pay} proposed \texttt{DetectVC} to detect voluntary followers (volowers). They showed how volowers gain profit by following enough users for self and product promotion. \textit{However, do these users act as a group to perform malicious activities?}  To answer this, \citet{gupta2019malreg} proposed approaches to detect malicious retweeter groups. They found that the activities appear normal at the individual level. However, at the group level, they look suspicious and group activities vary across groups. \citet{jang2019distance} is the first work that detects collusive followers using geographical distance. The authors reported that when the distance between the legitimate users and collusive users increases, the legitimate users' follower ratio between the number of followers and the total number of followers decreases. \citet{aggarwal2018follower} detected users on Twitter with manipulated follower count using an unsupervised local neighborhood detection method. 

    Other studies have looked at how collusion happens on Facebook and e-commerce platforms. \citet{de2014paying} conducted an in-depth analysis of Facebook pages where they collected likes via Facebook ads and other farms. The authors monitored the ``liking" activity by creating honeypot pages and crawling them after every 2 hours to check for new likes from the blackmarket services. \citet{badri2016uncovering} conducted an in-depth analysis of fake likers on Facebook collected from the blackmarket services. They reported that fake likers behave very differently than genuine likers in terms of their liking behaviors, longevity, etc. \citet{mukherjee2012spotting} proposed an unsupervised approach to detect and rank collusive spammer groups in Amazon.  They used Frequent Itemset Mining (FIM) to extract the set of reviewer groups. The extracted reviewer groups were labeled by the domain experts into spammer and genuine groups. 
% They also found that labeling reviewer groups are relatively easy compared to labeling individual reviewers. 
Their proposed method \texttt{GSRank} considers various group-level spam behavioral indicators, individual spam behavioral indicators and then models the inter-relationship between reviewer groups, members of those reviewer groups and products they reviewed to find the spammer groups. %They evaluated the resulting spammer groups against the labeled ones obtained earlier.

\begin{figure}[!t]
    \centering
    \subfloat[]{{\frame{\includegraphics[width=4.8cm]{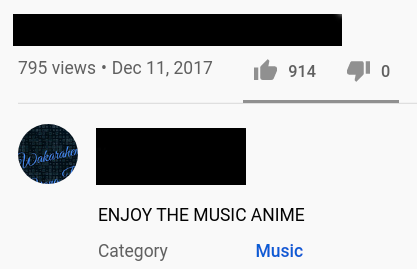}} }}%
    \subfloat[]{{\frame{\includegraphics[width=5.4cm]{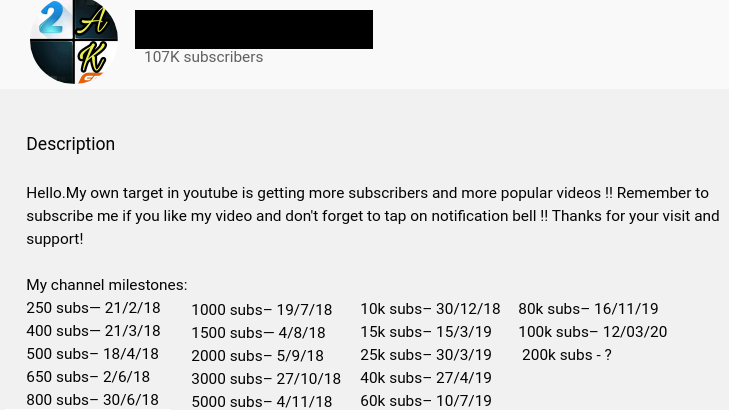} }}}%
    \hspace{0.26em}%
    \subfloat[]{{\frame{\includegraphics[width=5.2cm]{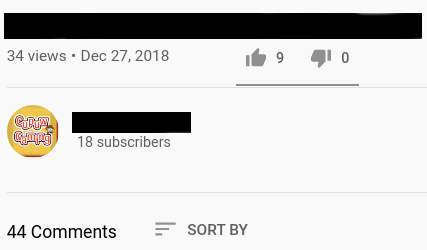}} }}%
    \caption{Example of (a) collusive video (for likes), (b) collusive channel (for subscriptions), and (c) collusive video (for comments) detected by our models. Sensitive information are blurred. Reprinted with permission from \cite{dutta2020detecting}.}%
    \label{fig:model_examples}%
\end{figure}

 Studies have also looked at artificial manipulation of viewership/comment/subscription count in the video streaming platforms such as YouTube, Twitch, etc. \citet{shah2017flock} introduced \texttt{FLOCK}, an unsupervised multi-step process for identifying botted views and botted broadcasts. \texttt{FLOCK} performs the following steps:  (a) it models a normal broadcast as a collection of multiple views; (b) it classifies whether a given broadcast has inflated popularity  by looking at the collective behavior of all the views in a broadcast;  (c) it classifies whether individual views of the broadcast are botted. For modeling the normal broadcast behavior, the authors included temporal features of the constituent views of broadcast, such as start time and end time of a particular view. It is also worth noting that they  avoided using descriptive features (like browser used, country of origin) and engagement-based features to make the proposed solution easier. In a very recent study, \citet{dutta2020detecting} proposed three models to detect three types of collusive entities on YouTube: (a) videos submitted for collusive like requests, (b) videos submitted for collusive comment requests, and (c) channels submitted for collusive subscription requests. They analyzed the videos submitted to blackmarket services for collusive appraisals (likes and comments) based on two perspectives -- propagation dynamics and video metadata. They also analyzed the collusive YouTube channels based on location, channel metadata, and network properties. Their analysis on the structural properties of the giant component present in the collusive channel network shows that it is a small-world. In order to detect the collusive videos and collusive channels, they designed one-class classifiers based on features calculated from video metadata and temporal information. The models were evaluated in terms of true positive rate, achieving state-of-the-art results. The authors also proposed \texttt{CollATe}, a denoising autoencoder model that leverages the power of three components: metadata feature extractor, anomaly feature extractor and comment feature extractor to learn feature representation of videos. Note that \texttt{CollATe} can only detect videos submitted in blackmarket services for collusive comment requests as unlike temporal information of like activity, the temporal information of the comments is publicly available in the YouTube API\footnote{\url{https://developers.google.com/youtube/v3}}. Fig. \ref{fig:model_examples} shows the snapshot of collusive entities detected using their model. Fig. \ref{fig:model_examples}(a) shows a video where the number of likes is much higher than the number of views. The authors claimed that the blackmarket services use YouTube API using the credentials of its users to help these videos  gain likes. Similarly, Fig. \ref{fig:model_examples}(c)
 shows the snapshot of a video submitted to blackmarket services for collusive comment subscriptions with a higher number of comments as compared to views and likes. Fig. \ref{fig:model_examples}(b) shows a YouTube channel that was submitted for collusive subscriptions. The authors claimed that collusive YouTube channels make proper use of promotional keywords to attract more subscribers to their channel.
 
The work by \citet{vidros2017automatic} is the only work that deals with recruitment based fraud detection. The authors introduced this problem as \textit{Online Recruitment Fraud} (ORF), more specifically relating to employment scams. The dataset used in this work contains numerous job ads and their corresponding annotations (legitimate or fraudulent). They performed an in-depth analysis of the dataset by applying standard state-of-the-art machine learning classifiers. In these cases, the trapped user(s) unknowingly become a medium for scammers to complete their jobs, often even driving the individual to complete a direct wire transfer for them under the careful and believable disguise of working visa or travel expenses. With this much being said about employment scams, it is worth noting that although they share some similar characteristics with problems such as email phishing, online opinion fraud, trolling, cyber-bullying, they are very much different in many aspects.

\textit{Limitation.} The main limitation is that in cases where user features are used, it can be restrictive to a particular online media platform and not generalize to others. Moreover, if collusive activities are performed by new online media accounts, it is difficult to collect user behaviors for such accounts. The drawback of using linguistic features is the lack of generalizability across multiple languages and circumstances. It has been reported in the literature \cite{ali2008language,larcker2012detecting} that linguistic features designed for one circumstance may not work perfectly for other circumstances.

\subsection{Graph Based Models}
In this section, we present and discuss the graph-based models for collusive entity detection. Recently, graph-based models have been developed particularly for spotting outliers and anomalies in networks. The advantage of using graph techniques is to efficiently capture long-range correlations among the inter-dependent data objects \cite{akoglu2015graph}.

\citet{Chetan:2019:CRD:3289600.3291010} proposed \corerank, an unsupervised approach to capture the interdependency between the \textit{credibility of users} and the \textit{merit of tweets} using a recurrence formulation based on network, behavioral information of users and topical diversity of tweets. The authors created a directed bipartite graph by modeling the interactions between users and
tweets in terms of retweets/quotes. They finally reported suspiciousness scores for both users and tweets in the graph based on the recurrent formulation. By incorporating prior knowledge about the collusive users,  they proposed a semi-supervised  version of \corerank, called \corerank+. Both \corerank~and \corerank+ outperformed their respective state-of-the-art methods by a significant margin.

\citet{rayana2015collective} proposed \texttt{SpEagle}, a framework that considers metadata of reviews (review content, rating, timestamp) along with relational data (reviewer-review-product graph) for detecting spam users, fake reviews and the products targeted by spammers. \texttt{SpEagle} used metadata for extracting the spam features, which were then transformed into a spam score to decide the genuineness of users and reviews. It was also extended to a semi-supervised setting (called \texttt{SpEagle+}) to improve its performance. The authors also implemented a lighter version (\texttt{SpLite})  that leverages a small subset of review features for the detection task to achieve a boost in runtime. \citet{wang2016detecting} proposed a bipartite graph projection-based approach, called \texttt{GSBP} to detect review spammer groups in e-commerce websites. Unlike FIM based methods \cite{mukherjee2012spotting}, their proposed approach detects loosely connected spammer groups -- each group member may not review every target product of the group. 
% Such type of spammer groups is more frequent in the real world. The authors proposed several group spam indicators to measure the spamicity of each group and then identified highly suspicious groups using a divide and conquer approach. 

\citet{kumar2018rev2} developed a fraudulent reviewer detection system, called \texttt{REV2} for rating platforms. They proposed three interdependent metrics -- {\em fairness} of a user in rating an item,  {\em reliability} of a specific rating, and {\em goodness} of a product. By combining network and behavioral properties of users, products and ratings, \texttt{REV2} calculates fairness, reliability, goodness scores in both supervised and unsupervised settings. They evaluated the computed scores by using five real-world datasets -- Flipkart, Bitcoin OTC, Bitcoin Alpha, Epinions and Amazon. The current version of \texttt{REV2} is deployed at Flipkart (an Indian e-commerce company). 

\citet{wang2018graph} proposed a graph-based framework, called \texttt{GGSpam} to detect  review spammer groups. It recursively splits the entire reviewer graph into small  groups. \texttt{GSBC}, the key component of \texttt{GGSpam}, finds the spammer groups by identifying all the bi-connected components within the split reviewer graphs whose spamicity score exceeds the given threshold. They proposed several group spam indicators to compute the spamicity score of a given reviewer group. They conducted experiments on two real-world datasets -- Amazon and Yelp. 

\begin{figure}[!t]
    \centering
    \subfloat[]{\includegraphics[width=0.33\columnwidth]{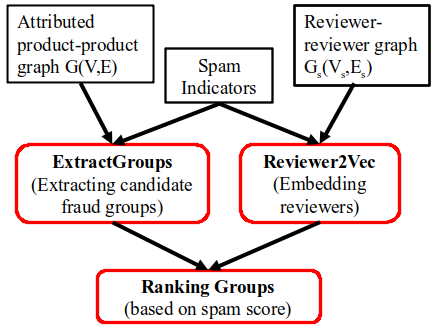} }%
    \hspace{0.2mm}
    \subfloat[]{\includegraphics[width=0.64\columnwidth]{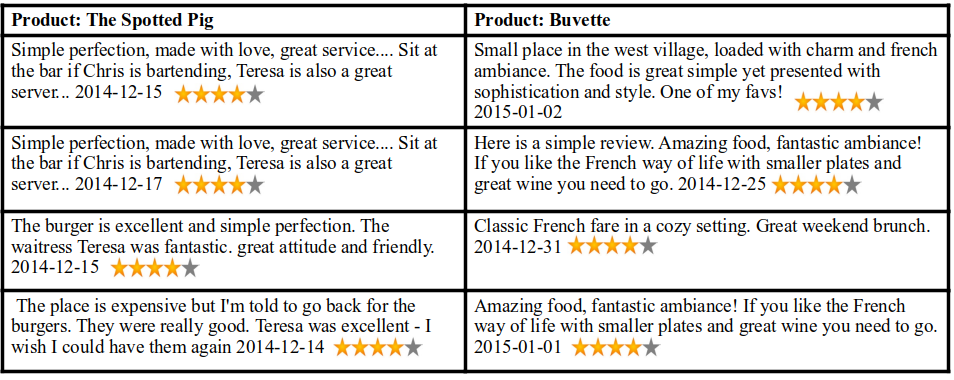} }%
    \caption{(a) Schematic diagram of DeFrauder to detect online fraud reviewer groups. (b) Coherent review patterns of four spammers in a fraudulent reviewer group.  Reprinted with permission from \cite{dhawan2017spotting}. }
    \label{fig:defrauder}
\end{figure}

\citet{dhawan2017spotting} proposed \texttt{DeFrauder}, an unsupervised approach to detect online fraud reviewer groups. This approach initially detects the candidate fraud groups by leveraging the underlying product review graph and incorporating several behavioral signals which model multi-faceted collaboration among reviewers. It then maps the reviewers into an embedding space and assigns a spam score to each group, such that groups consisting of spammers with highly similar behavioral traits attain high spam score. They conducted experiments on four real-world datasets -- Amazon, PlayStore, YelpNYC \cite{rayana2015collective} and YelpZip \cite{rayana2016collective}. Fig. \ref{fig:defrauder}(a) shows the schematic diagram of DeFrauder to detect online fraud reviewer groups. Fig. \ref{fig:defrauder}(b) shows the coherent review patterns of four spammers in a fraudulent reviewer group.

\citet{liu2017holoscope} formulated a  novel suspiciousness metric from graph topology and spikes to detect fraudulent users in multiple rating platforms (BookAdvocate, Amazon). The authors proposed \texttt{HoloScope}, an unsupervised approach that combines suspicious signals from graph topology, temporal bursts and drops, and rating deviation. The primary goal behind the approach is to obstruct the fraudsters by increasing the time they need to perform an
attack. The evaluation was done on multiple semi-real (synthetic) and real datasets.

\citet{shin2018fast} proposed two algorithms -- \texttt{M-Zoom} (Multi-dimensional Zoom) and \texttt{M-Biz} (Multidimensional Bi-directional Zoom) to detect suspicious lockstep behavior in online review platforms. The task of detecting lockstep behavior is to find out a set of accounts that give fake reviews to the same set of products/restaurants. The objective of \texttt{M-Zoom} and \texttt{M-Biz} is to detect dense subtensors in a greedy way until it reaches a local optimum. Though the overall structure of \texttt{M-Biz} and \texttt{M-Zoom} is the same, it differs significantly in the way they find each dense subtensor. Using the above algorithms, the authors detected three dense subtensors that indicated the activities of bots, which changed the same pages hundreds of thousands of times.

\citet{mehrotra2016detection} detected fake followers on Twitter using graph centrality measures. The authors created a graph using the follower and friend network from a set of Twitter users. Next, six centrality-based features were used for the classification experiment. Using this feature set with a Random Forest classifier gave an accuracy of $95\%$. \citet{zhang2016discover} proposed a graph-based approach using near-duplicates to detect fake followers in Weibo, a Chinese counterpart of Twitter. They were able to discover 11.90 million users in Weibo who bought followers from blackmarket services.

\citet{shin2016corescope} detected three empirical patterns of core users in real-world networks across diverse domains. The first pattern is the \texttt{MIRROR PATTERN}, which states that the coreness of a vertex is strongly related to its degree. The second pattern \texttt{CORE-TRIANGLE PATTERN} states that the degeneracy (non-core component) and the triangle-count obey a power-law with slope 1\/3. The third pattern \texttt{STRUCTURED CORE PATTERN} states that the degeneracy cores are not cliques. Using these patterns, the authors analyzed the lockstep behavior for collusive followers on Twitter. It was reported that at least 78\% of the vertices with high coreness values  use blackmarket services to boost their followers.

\textit{Limitation.} Graph-based methods are computationally intensive which can restrict their ability to design real-time tools to detect collusive entities in online media platforms.

\subsection{Deep Learning Based models}
The advent of deep learning has alleviated the shortcomings of the previous models with the power to learn more complex representations that are difficult to be captured by traditional machine learning models.

\citet{arora2019multitask} proposed a multi-task learning approach to detect collusive tweets. The authors collected the data from two blackmarket services - YouLikeHits and Like4Like after creating honeypot accounts in these services. The multi-task model takes two inputs -- the feature representation extracted from the tweet metadata and Tweet2vec representation of the tweets, which are passed through a cross-stitch neural network to learn the optimal combination of the inputs via soft parameter sharing. \citet{ahsan2016review} tackled the problem of fake reviewers using an active learning approach with the TF-IDF features of the review content.

Studies have also looked at the detection of social spammers in online media platforms. \citet{wu2020graph} proposed an end-to-end deep learning model based on Graph Convolution Networks (GCNs) that operates on directed social graphs to detect social spammers. The authors built a classifier by performing graph convolution on the social graph with different types of neighbors.  

In a study to identify collusive entities on YouTube, \citet{dutta2020detecting} proposed \texttt{CollATe}, a deep unsupervised learning method using denoising autoencoders. The \textit{CollATe} architecture consists of three components -- metadata feature extractor, anomaly feature extractor, and comment feature
extractor to learn video feature representations. The final representation is then fed to a collusive video detector module that returns the label -- \textit{collusive} or \textit{non-collusive} for the input video.

To detect collusive followers on Twitter, \citet{castellini2017fake} proposed a deep learning technique using a denoising autoencoder. The denoising autoencoder is implemented as an anomaly extractor and is trained with examples only from the features generated from the real profiles. The authors built a test set with both real and fake profiles. Once they got the reconstruction error calculated using the autoencoder for a data point, it was then compared to a threshold value (computed from the training set): if the error is higher, it is marked as collusive (i.e., the autoencoder is not able to properly reconstruct the record), otherwise, it is marked as real.

\textit{Limitation.} The main limitation of the deep learning methods is the requirement of vast amounts of human-annotated data. Moreover, due to the unusual behavior of collusive entities due to the mixture of organic and inorganic activities, it is difficult to find correlations between sets of features which makes the deep learning techniques to not perform well.

\subsection{Other Methods}
In contrast to the above approaches which were largely focused on collusive activities in online media platforms, a series of works have been conducted on detecting collusive activities in other platforms. 

\citet{asavoae2016towards} developed a fully automated and effective detection system to detect colluding android apps that violate the permissions causing data leaks or distributing malware over multiple apps. The authors first proposed a rule-based approach to identify colluding apps. As defining rules requires expert knowledge and for some cases explicit rules may not exist, the authors proposed a probabilistic model to calculate two likelihood components for an app: (i) the likelihood of carrying out a threat and (ii) the likelihood of performing an inter-app communication. 

\citet{blasco2018automated} proposed \texttt{Application Collusion Engine (ACE)}, a system to automatically generate colluding apps with a variety depending on the configuration of app component templates and code snippets. The \texttt{ACE} system has two main components:  \textit{colluding set engine} and \textit{application engine}. The first component tells the second component how it should create apps in order to collude. The second component is responsible for creating fully working android apps that are ready to be installed in a device. The primary aim of this work is to create substantial colluding app sets for experimentation as representative datasets do not presently exist for colluding apps. \citet{kalutarage2017towards} proposed a probabilistic model towards an automated threat intelligence system for app collusion. The readers are encouraged to go through \cite{bhandari2017android} for a detailed survey on the app detection techniques that focus on collusive activities by investigating inter-app communication threats.

\section{Annotation Guidelines, Datasets, Applications and Evaluation metrics} \label{sec:datasets}
As collusive entity detection is a considerably new research area, most of the previous studies suggested few guidelines which may be helpful to label these users. In this section, we first review the annotation guidelines for creating  annotated datasets for collusive entities. We then provide pointers to the previous studies which have released various datasets for collusive entity detection. Next, we discuss the tools and interfaces in the analysis and detection of collusive entities in online media. Finally, we describe the common evaluation metrics used for collusive entity detection.
\subsection{Annotation Guidelines}
Annotation guidelines are usually seen in studies that create a labeled set of entities for various experiments, as mentioned below.

The dataset of \citet{dutta2018retweet} is one of the most widely used dataset \cite{dutta2019blackmarket,arora2020analyzing} on collusive retweeter detection. The dataset is created from the blackmarket services and the annotators were asked to label each retweeter by one of the three classes: \textit{Bots}, \textit{Promotional Customers}, \textit{Normal Customers}. For instance,  \textit{Bots} can be selected if the account is controlled by software. The \textit{promotional customers} can be identified if the accounts are involved in promoting brands using keywords such as `win', `ad', `Giveaway'. The \textit{normal customers} class can be selected if the account does not fall under any of the above categories. The annotators were also given complete freedom to search for any information related to collusive entities on the web and apply their own intuition.

In \cite{xu2013uncovering}, a corpus of collusive spammers in a Chinese review website is described. The task of the annotator is to label each reviewer in a reviewer group as either colluder or non-colluder. The annotators in this work were computer science major graduate students who had extensive online shopping experiences. The annotators were given a set of three instructions about the colluder class and how they should reason to arrive at the colluder class when the data is not straightforward: (i) if the reviewer has too many reviews giving opposite opinions than other reviews about the same stores, (ii) if the reviewer has too many reviews giving opposite opinions about the same stores as compared to the ratings from Better Business Bureaus\footnote{\url{https://www.bbb.org/}}, an organization that focuses on advancing marketplace trust, and (iii) if the reviewer has too many reviews giving opposite opinions about the same stores as compared to evidence by general web search results.  

In \cite{mukherjee2013spotting}, a corpus of collusive spammers in an e-commerce website is described. Three annotators were given the task of examining the e-commerce users and provide a label as spammer or non-spammer based on three opinion spam signals. Note that each of the annotators was domain expert, i.e., an employee of e-commerce platforms such as Rediff Shopping, eBay.in, etc. As the annotators were domain experts, they were also given access to detailed information, e.g., entire profile history, demographic information, etc.

\citet{li2011learning} built a review spam corpus. They crawled data from a consumer review site, named Epinions\footnote{\url{http://www.epinions.com/}}. The task of the annotator was to annotate the review spam dataset and label each review as spam or non-spam. The annotators used the rules specified in a platform, named \texttt{Consumerist}\footnote{\url{https://consumerist.com/}}, an independent source of consumer news and information published by Consumer Reports, to label the reviews. The platform suggests a set of $30$ rules which are helpful for annotators to spot fake online reviews. The authors employed ten college students to annotate their reviews. Each review was annotated by two students, and conflict was resolved by another student. 

In most of the above studies, Fleiss'/Cohen's kappa value was reported to show the reliability of agreement between the annotators.

\subsection{Datasets}
In this section, we present a list of publicly-available datasets used for the detection of collusive activities in online media. 
%Though there is a lack of large-scale publicly available datasets for this problem, we tried our best to provide a detailed description of the available ones with their comparisons. 
Table \ref{table:comparison_datasets} compares these datasets and lists the available entities, as mentioned in their respective papers. 
% We first present the datasets related to OSNs, and then the ones related to rating and other platforms. 

\begin{table*}[!t]
\caption{A summary of the publicly available datasets of collusive entities in online media.}\label{table:comparison_datasets}

\scalebox{0.9}{
\begin{tabular}{|l|l|l|l|l|}
\hline
\multicolumn{1}{|c|}{\bf Name} &  \multicolumn{1}{c|}{\bf Available entities}  & \multicolumn{1}{c|}{\bf Type} & \multicolumn{1}{c|}{\bf Platform} & \multicolumn{1}{c|}{\bf Ref.}\\\hline
\score & Collusive users registered in freemium blackmarket services &  Individual  & Twitter & \cite{dutta2018retweet} \\
\corerank & Collusive users and tweets registered in blackmarket services &  Individual & Twitter & \cite{Chetan:2019:CRD:3289600.3291010}\\
\hawkeseye & Collusive retweeters promoting popular hashtags & Individual  & Twitter & \cite{dutta2020hawkeseye} \\
\mtlcollu & Collusive tweets registered in freemium blackmarket services  & Individual  & Twitter & \cite{arora2019multitask} \\
\scoreplus & Collusive users registered in freemium/premium blackmarket services &  Individual & Twitter & \cite{dutta2019blackmarket} \\
\fameforsale & Users involved in blackmarket-based following activities&  Individual  & Twitter & \cite{cresci2015fame}\\
\viewwgcca & Collusive users registered in freemium blackmarket services  & Individual & Twitter & \cite{arora2020analyzing} \\
\texttt{MalReg} & Malicious retweeter groups  & Group &Twitter & \cite{gupta2019malreg} \\
\texttt{DetectVC} & Voluntary following activities  & Individual & Sina Weibo & \cite{liu2016pay} \\
\texttt{FakeLikers} & Fake likers on Facebook  & Individual & Facebook & \cite{badri2016uncovering}\\
\gsrank & Review spammer groups &  Individual & Amazon & \cite{mukherjee2012spotting}\\
\gsbp & Loosely connected review spammer groups &  Individual & Amazon & \cite{wang2016detecting}\\
\revtwo & Fraudulent users in rating platforms & Individual & Amazon & \cite{kumar2018rev2}\\
\ggspam & Review spammer groups modeled as bi-connected components &  Group  & Amazon & \cite{wang2018graph}\\
\speagle & Spam users, fake reviews and the targeted products&  Individual  & Yelp & \cite{rayana2015collective}\\
\defrauder & Fraud reviewer groups &  Group & PlayStore & \cite{dhawan2017spotting}\\ 
\emscad & Online recruitment frauds & Individual  & LinkedIn & \cite{vidros2017automatic}\\
\texttt{CollATe} & Collusive videos and channels on YouTube &  Individual & YouTube & \cite{dutta2020detecting}\\

\hline

\end{tabular}}
\end{table*}
The \fameforsale~dataset \cite{cresci2015fame} is a corpus of fake followers collected from various blackmarket services -- InterTwitter (INT), FastFollowerz (FSF) and TwitterTechnology (TWT). The dataset also contains the relationships between the user accounts (followers/friends). The \score~dataset \cite{dutta2018retweet,dutta2019blackmarket} is a medium-sized corpus of collusive users, collected from four freemium blackmarket retweeting services -- YouLikeHits, Like4Like, Traffup and JustRetweet. It contains an anonymized version of $1,941$ users, who were manually annotated into four categories (bots, promotional customers, normal customers and genuine users). 
The authors also reported the values of $64$ features for each user, which were used in their approach to distinguish between collusive users and genuine users.
The \corerank~dataset \cite{Chetan:2019:CRD:3289600.3291010} is another corpus of retweeters collected from two blackmarket services -- YouLikeHits and Like4Like. It contains the details of $4,732$ collusive users and $2,719$ genuine users. \citet{liu2016pay} released the \texttt{DetectVC} dataset of voluntary followers containing $3,250$ volower IDs and their following relations. The \texttt{MalReg} dataset \cite{gupta2019malreg} contains annotated groups of users that collude to retweet together maliciously. It consists of $1,017$ malicious retweeter groups collected from three political events: UK General Election 2017, Indian banknote demonetization 2016 and Delhi legislative assembly election 2013. The \texttt{FakeLikers} dataset contains $13,147$ likers, $4.66$M pages, $0.99$M posts and $5.47$M friend relations from Facebook. There is only one available dataset for recruitment frauds (\emscad) \cite{vidros2017automatic}. The dataset consists of $17,880$ annotated job ads ($17,014$ legitimate and $866$ fraudulent) published between 2012 and 2014 and annotated by people having domain expertise. Each job ad is described by a set of fields, and a label indicates whether the entry is fraudulent or not. There are a few publicly available datasets for rating platforms. \citet{dhawan2017spotting} released a  large-scale dataset of reviews from different applications
available on Google Playstore. The dataset contains $3,25,424$ reviews made by $3,21,436$ reviews on $192$ apps. YelpNYC \cite{rayana2015collective} and YelpZip \cite{rayana2016collective} are the datasets collected  from Yelp, and used for detecting collusive users/groups. YelpNYC dataset consists of review data related to $923$ restaurants in New York City. It includes $359$k reviews and $160$k reviewers. YelpZip is relatively large compared to the other two datasets. It consists of $608$k reviews and $260$k reviewers, for $5$k restaurants located in multiple states of the United States. \citet{dutta2020detecting} released their dataset on collusive YouTube entities. The dataset contains a large set of YouTube videos ($45,572$ videos for like appraisals, $25,106$ videos for comment appraisals) and channels ($7,847$ channels for subscription appraisals) submitted to the blackmarket services for collusive appraisals.

\subsection{Interfaces and Applications}
\begin{figure}[!htbp]
    \centering
    \subfloat[]{\includegraphics[width=0.49\columnwidth]{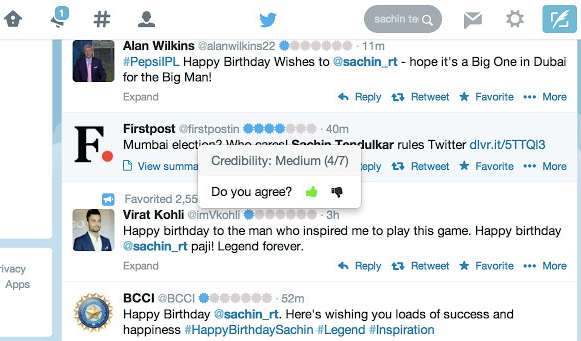} }%
    \hspace{0.2mm}
    \subfloat[]{\includegraphics[width=0.47\columnwidth]{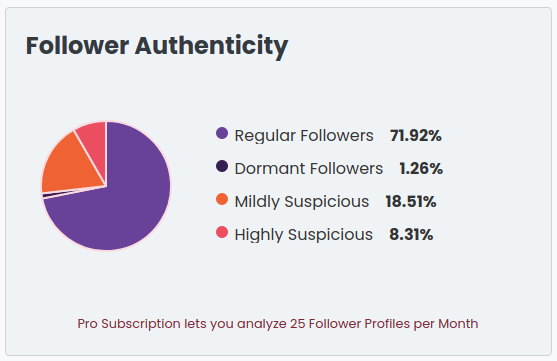} }%
    \caption{(a) Working of \texttt{TweetCred} to assess credibility of content on Twitter. (b) Working of \texttt{Analisa} to find the authenticity of followers on Instagram. }
    \label{fig:example_applications}
\end{figure}

In this section, we discuss the tools and interfaces developed to detect collusion in online media. A list of interfaces and applications is provided below with references to the corresponding studies as well. Fig. \ref{fig:example_applications} shows the working of two such applications. 
\begin{itemize}
    \item \citet{dutta2018retweet} developed a chrome extension, called \score~ to detect collusive users involved in blackmarket-based retweeting activities. \score~also supports online learning by incorporating user feedback.
    \item \texttt{TwitterAudit}\footnote{\url{https://www.twitteraudit.com}} and \texttt{FollowerWonk}\footnote{\url{https://followerwonk.com/analyze}} services analyze Twitter accounts to check for fake followers.
    \item \texttt{Botometer}\footnote{\url{https://botometer.iuni.iu.edu}} assigns a score (0-5) to an account based on its bot activities. Higher the score, higher the chance that the account is being controlled by a bot.
    \item \texttt{TweetCred} \cite{gupta2014tweetcred} measures the credibility of a tweet based on  user-centric and tweet-centric properties.
    \item \texttt{FakeCheck}, \texttt{FakeLikes} and \texttt{IGAudit}\footnote{\url{https://www.fakecheck.co}, \url{https://fakelikes.info/}, \url{https://igaudit.io/}} evaluate Instagram accounts for fake followers.
    \item \texttt{TwitterPatrol} \cite{singh2018twitter} is a real-time tool to detect spammers, fake and compromised accounts on Twitter. 
    \item \texttt{Modash}\footnote{\url{https://www.modash.io/}} analyzes brands to find influencers on Instagram. It also has the functionality to detect fake followers on a public user account.
    \item \texttt{Analisa}\footnote{\url{https://analisa.io/}} is a tool for bloggers and agencies to check the follower authenticity for an Instagram or TikTok account.
\end{itemize}     

Finally, some of the previous works have publicly released their resources and codes. Interested readers can refer to the following studies: \cite{dutta2018retweet,dutta2019blackmarket,arora2019multitask,arora2020analyzing,dhawan2017spotting,gupta2019malreg} where public links to the resources and codes for collusive entity detection and analysis are available. We believe that such studies are worthy of attention as they encourage the reproducibility of the experiments and enable the follow-up studies to use the models as baselines to improve the collusive entity detection models.

\subsection{Evaluation Metrics}
For any classification task, metrics such as \textit{precision}, \textit{recall}, and \textit{F-score} are commonly used. However, in most of the studies, we observe \textit{F-score} to be more popular as it is a combined metric that conveys the balance between the \textit{precision} and \textit{recall}. In general, \textit{F-score} is calculated as the weighted harmonic mean of precision and recall, with a beta parameter ($\beta$) which determines the weight of recall. In collusive entity detection, the most widely used version of F-score is calculated as the macro-average of the F-scores for \textit{collusive} and \textit{genuine} classifications as follows:

\begin{align*}
 F &= \dfrac{2 \times precision \times recall}{Precision + Recall}  & F &= \dfrac{F_{collusive} + F_{genuine}}{2}  \\
\end{align*}
\begin{align*}
    F_{collusive} &= \dfrac{2 \times Precision_{collusive} \times Recall_{collusive}}{Precision_{collusive} + Recall_{collusive}}  & F_{genuine} &= \dfrac{2 \times Precision_{genuine} \times Recall_{genuine}}{Precision_{genuine} + Recall_{genuine}}  \\
\end{align*}
\begin{align*}
    Precision_{collusive} &= \dfrac{Correct_{collusive}}{Correct_{collusive} + Spurious_{collusive}}  & Precision_{genuine} &= \dfrac{Correct_{genuine}}{Correct_{genuine} + Spurious_{genuine}}  \\
\end{align*}
\begin{align*}
    Recall_{collusive} &= \dfrac{Correct_{collusive}}{Correct_{collusive} + Missing_{collusive}}  & Precision_{genuine} &= \dfrac{Correct_{genuine}}{Correct_{genuine} + Missing_{genuine}}  \\
\end{align*}
where $Correct_{collusive}$ and $Correct_{genuine}$ indicate correctly classified collusive and genuine users, respectively. Similarly, $Spurious_{collusive}$ ({\em resp.} $Missing_{collusive}$)  and $Spurious_{genuine}$ ({\em resp.} $Missing_{genuine}$) indicate false positive ({\em resp.} false negative) for collusive and genuine classes, respectively. 

\textit{Precision@k} and \textit{Recall@k} are another metrics used for collusive entity detection \cite{Chetan:2019:CRD:3289600.3291010} and are calculated as follows:

\begin{align*}
 Precision@k &= \frac{\text{\# of recommended items @k that are collusive}}{\text{k}}  \\
 Recall@k &= \frac{\text{\# of recommended items @k that are collusive}}{\text{total \# of collusive items}}.
 \end{align*}
 
%Most of the previous studies evaluated the performance of their methods using the above metrics.
\citet{dutta2020detecting} also reported the True Positive Rate (TPR). The reason behind choosing TPR as the evaluation metric is because all their models are trained on one-class, and the authors are only interested in the proportion of actual positives (data in the \textit{collusive} class) that are correctly identified by the models. The evaluation strategy can be considered when the objective is to measure the effectiveness of a single class.

\section{Open Problems} \label{sec:future_opportunities}
Collusion is a considerably new topic in the area of anomaly detection which opens up a number of future research opportunities. Some of these topics include: (i) Collective collusion detection, (ii) understanding connectivity patterns in collusive network, (iii) event-specific studies, (iv) temporal modeling of collusive entities, (v) cross-lingual and multi-lingual studies, (vi) core collusive user detection, (vii) cross-platform spread of collusive content, (viii) multi-modal analysis of collusive entities, and (ix) how collusion promotes fake news. We briefly discuss these topics in detail below.

\subsection{Collective Collusion Detection} Recently, collusive entity detection has gained a lot of attention in the literature. However, most of the existing methods only detect independent collusion; but in reality, it is often seen that the anomalous phenomena also happens in groups. The primary aim of the collusive group detection task is to find a set of users that jointly exhibit anomalous behavior. Detecting anomalous groups is more difficult as compared to the individual detection task due to the inter-group dynamics. We request the reader to read Section \ref{sec:collusion_types} for details of previous studies in individual and group collusive entity detection. We believe that the advent of new datasets (see Section \ref{sec:datasets} for more details) will foster fraudulent user/entity detection (both individual and group level) with the advantage of adding the topical as well as the temporal dimension. 

\subsection{Understanding Connectivity Patterns in Collusive Network} Understanding connectivity patterns of an underlying network is a well-studied problem in the literature.  It includes tasks such as inferring lockstep behavior \cite{beutel2013copycatch}, dense block detection \cite{shin2016m}, detecting core users \cite{shin2016corescope,shin2018patterns}, identifying the most relevant actors in a network \cite{borgatti2006identifying}, sudden  appearance/disappearance of links \cite{eswaran2018spotlight}, etc. One potential research direction is to create various networks among the entities to investigate the network's structural patterns. Recently, several studies have  modeled  information diffusion for collusive entities from a topological point of view. It includes tasks such as influence maximization (selecting a seed set to maximize the influence spread) \cite{jendoubi2017two,mei2017influence}, predicting information cascade \cite{rattanaritnont2011study,hakim2014predicting}, measuring message propagation and social influence \cite{ye2010measuring,brown2011measuring}, etc.

\subsection{Event-specific Studies} Event-specific studies can be employed by deeply investigating large-scale datasets. Some of the publicly available datasets mentioned in Section \ref{sec:datasets} are obtained from multiple sources and span over a long period of time. Therefore, it may consist of information from many major events \cite{atefeh2015survey}, which can be easily extracted for event-centric studies. Researchers can also check how these users/entities were involved in manipulating the popularity of events by artificially inflating/deflating the social growth of entities in online media \cite{zhang2016twitter}.

\subsection{Temporal Modeling of Collusive Entities} A topic closely related to the previous topic is the temporal modeling of the collusive entities. The tasks in temporal modeling include detecting time periods containing unusual activity\cite{giatsoglou2015retweeting}, identifying repetitive patterns in time-evolving graphs \cite{zeidanloo2010botnet}, etc. Recently, detecting anomalies in streams \cite{eswaran2018spotlight,eswaran2018sedanspot} has gained a lot of attraction in the research community due to the time-evolving (or dynamic) property where consecutive snapshots of activity in a time window are monitored. Related studies for temporal modeling can be found in \cite{wilmet2018degree,yoon2019fast,eswaran2020mining}.

\subsection{Cross-lingual and Multi-lingual Studies} The datasets in collusive entity detection are collected from various online media platforms where the entities usually have texts written in several languages. Though we observe the lack of annotated datasets in languages other than English, cross-lingual studies can be done by converting the English texts into the target language using automated translation tools. The converted texts can then be used to create the training and test datasets. For multi-lingual studies, we can consider it as a future research topic that can be performed only when sufficient content is available in multiple languages.

\subsection{Core Collusive User Detection}
The underlying blackmarket collusive network comprising two types of users: (i) {\em core users} -- fake accounts or sockpuppets,  which are fully controlled by the blackmarkets (puppet masters), and {\em compromised accounts} that are temporarily hired to support the core users.  These two types of users are together called as {\em collusive users}. Core users are the spine of any collusive blackmarket; they  monitor and intelligently control the entire fraudulent activities in such a way that none of their hired compromised accounts are being suspended. Therefore, detecting and removing core blackmarket users  is of the utmost importance to decentralize the collusive network and keep the YouTube ecosystem healthy and trustworthy.

According to the literature \cite{huang2020identifying}, detecting core nodes in a network is to find the influential nodes. However, in the case of collusion, core nodes might not be influential as these are the accounts that are fully controlled by the blackmarket authorities. The work by \citet{shin2016corescope} is the only study  that explores empirical patterns of core users in real-world networks. \cite{huang2020identifying,zhang2017top} are some of the studies on influential node detection in a network. However, none of the existing studies attempted to detect core collusive users. One reason may be the lack of ground-truth for training the model and evaluation. We believe that Core blackmarket user detection is highly important to understand how these services operate and flag their behavior. 

\subsection{Cross-platform Spread of Collusive Content}
Another future work could be to investigate the cross-platform spread of collusive entities and study its impact. Online media platforms differ from each other in multiple ways: (i) different platforms have significantly different language characteristics, (ii) some platforms have restrictions on the length of posts allowing users to express themselves within less space, (iii) some platforms only allow images/videos as posts, and (iv) different platforms have different types of appraisals. Very few studies considered analyzing cross-platform data \cite{samani2018cross,jaidka2018facebook} in online media platforms. Moreover, no work to date examined the cross-platform study of collusive entities. An important related issue to conduct the cross-platform study is the need for ground-truth datasets collected from different online media platforms. 

\subsection{Multi-modal Analysis of Collusive Entities}
A recent trend in social computing research is to conduct a multi-modal analysis of online media entities. The multi-modal investigation of collusive entities is important due to the following reasons: (i) different modalities may exhibit different but important information about an entity (e.g., may help in detecting the authenticity of the information), and (ii) different modalities can be manipulated differently by the blackmarket services. Note that it is expected that multi-modal analysis will introduce a computational cost due to the operation on image and video data.  \citet{singhal2019spotfake} proposed a multi-modal framework, called \texttt{SpotFake} for fake news detection by leveraging the textual and image modalities. Hence, new models can be designed to  incorporate different modalities for collusive entity detection.

\subsection{How Collusion Promotes Fake News}

Online social networks  are a popular way of dissemination and consumption of information.
However, due to their decentralized nature, they also come with limited liability for disinformation and
collusive persuasion. It is often the experience that online users are subjected to a barrage of misinformation
within a short span of time, all of which are targeted at persuading the user to develop a particular sentiment
against a person, a political party, a system of medicine, or the cause of an event. This ``echo chamber'' effect
influences users' thoughts in a manner that is unfavourable to the spirit of free speech and debate. This direction may undertake research into the detection of fake news within online social networks, with a
specific focus on understanding the nature of collusive orchestration of misinformation. In particular, one may consider -- (i) collusion in fake news, i.e.,  identifying tell-tale patterns of
collusion within the context of fake news, and (iii) collusive fake news detection, i.e., leveraging such collusive
patterns in order to enhance fake news detection. In addition to improving fake news detection, this project will make fundamental advances towards a novel
direction of research in fake news analytics; that of characterizing and exploiting collusive behavior towards
improving fake news detection.

%Fake news can be seen as a specific type of disinformation that is usually spread through online media platforms using text posts, video posts, memes, unverified advertisements and rumors. We request the reader to refer \cite{zhou2018fake} for a detailed survey on fake news. The nature and characteristics of collusion differ from fake news in multiple aspects (see Table \ref{table:comparison_definition} for more details); hence the detection techniques of fake news are not applicable to collusion. However, it will be interesting to see how appraisals in fake news are artificially manipulated with the help of blackmarket services. A recent study conducted by \cite{shao2017spread} shows how fake news is propagated by bots and fake accounts in online media platforms. 

\section{Conclusion and Open Challenges} \label{sec:conclusion}
In this survey paper, we presented a detailed overview of the fundamentals of collusion and detection of artificial boosting of social growth and manipulation observed in online media. We categorized existing studies based on the platforms which are compromised by the collusive users. The approaches were mostly developed for the detection of collusive entities using techniques such as feature-based, graph-based and deep learning-based in both supervised and unsupervised settings. In addition, the survey also includes the related annotation guidelines, datasets, available applications, evaluation metrics and future research opportunities for collusive entity detection.  

This area is still in its infancy and has various open challenges. The major challenge is the collection of large-scale datasets, as crawling blackmarket services is extremely challenging and may require ethical consent. Moreover, there are restrictions and limitations of the APIs provided by the online media platforms. Furthermore, human annotation to label an action (retweet, share, like, etc.) as collusive is confusing and often incurs low inter-annotator agreement \cite{Chetan:2019:CRD:3289600.3291010} as collusive activities bear high resemblance to genuine activities. Therefore, it is difficult to design efficient supervised methods. There are limited studies on the detection of such collusive manipulation in online media platforms other than social networks, which requires detailed inspection. Detection and evaluation of self-collusion (such as creating multiple-account deception, sockpuppets, etc.) are extremely challenging due to the difficulty in collecting ground-truth data. We believe that this survey will motivate researchers to dig deeper into exploring the dynamics of collusion in online media platforms.  

\bibliographystyle{ACM-Reference-Format}
\bibliography{csur}

\end{document}